\documentclass[preview]{elsarticle}
\usepackage{graphicx}
\usepackage{float} 
\usepackage{epstopdf, epsfig}
\usepackage{amsmath} 
\usepackage{amssymb} 
\usepackage{physics} 
\usepackage{xcolor} 
\usepackage{hyperref} 
\usepackage{stmaryrd} 
\usepackage[normalem]{ulem} 
\usepackage[percent]{overpic} 
\usepackage{tikz} 
\usepackage{subcaption}  

\newcommand{\vect}[1]{\boldsymbol{#1}} 
\newcommand{\matr}[1]{\boldsymbol{\mathsf{#1}}}
\newcommand{\vecf}[2]{\vect{#1}(\vect{#2})} 

\newcommand{\nabl}{\vect{\nabla}}

\let\today\relax
\makeatletter
\def\ps@pprintTitle{%
    \let\@oddhead\@empty
    \let\@evenhead\@empty
    \def\@oddfoot{\footnotesize\itshape
         {} \hfill\today}%
    \let\@evenfoot\@oddfoot
    }
\makeatother

\begin{document}
\begin{frontmatter}

\title{A 3D BEM algorithm for simulations of magnetic fluid droplet dynamics}
\author[a]{Aigars Langins}
\ead{aigars.langins@lu.lv}
\author[a,b]{Andris P.Stikuts}
\author[a]{Andrejs Cēbers}

\address[a]{MMML Lab, Department of Physics, Mathematics and Optometry, University of Latvia, Jelgavas st. 3--014, Riga, LV-1002, Latvia }
\address[b]{Laboratoire PHENIX, Sorbonne Université, CNRS, 4 place Jussieu, case 51, F-75005 Paris, France}

\begin{abstract}

This paper outlines a numerical algorithm that could be used for simulating full 3D dynamics of magnetic fluid droplet shapes in external magnetic fields, by solving boundary integral equations. 
The algorithm works with arbitrary droplet and carrier fluid viscosity ratios. It is validated with known theoretical relationships.

Thus it may be used to evaluate various approximations often used in description of ellipsoidal droplets by comparing droplet dynamics calculated from them to the results obtained numerically from first principles here.

The algorithm may be used for investigations of droplet configurations in arbitrary magnetic fields, as well as for indirectly calculating its physical properties and predicting the magnetic field thresholds above which droplet shape can develop instabilities in the form of various spikes.
\end{abstract}


\end{frontmatter}

\section{Introduction} 

Figures of equilibrium of self-gravitating masses is a classical problem of mathematical physics \cite{chandrasekh_eqlb_fig_69}, where different bifurcations of shapes are described. 
Since the classical paper by G.I.Taylor \cite{taylor-disintegration-1964} the equilibrium shapes and their dynamics under the action of electromagnetic fields have been investigated in many works.
Significant breakthrough in the research of droplets under the action of electromagnetic field arose with the synthesis of magnetic liquids \cite{rosensweig_ferrohydrodynamics_1985} allowing many interesting effects to be observed and described, such as the droplet deformation and its dynamics under the action of static magnetic fields \cite{drozdova_79, bacri_instability_1982, osti_6426307}, rotating fields \cite{bacri_behavior_1994, morozov_bifurcations_2000} and labyrinthine pattern formation in the Hele-Shaw cells \cite{cebers80_interf_struct}.
For the description of these effects different approximate methods \cite{bacri_bistability_1983,cebers_virial_1985,afkhami-deformation-2010} (assumption of ellipsoidal shape, satisfaction of boundary conditions on average and others) were created which need to be confirmed.
Even more different observed phenomena still are not described theoretically or sufficiently explored numerically, -- such as the dynamics of hysteresis of droplet deformation \cite{bacri_dynamics_1983}, reentrant transition of figures of equilibrium of magnetic droplets in a high frequency rotating field \cite{bacri_behavior_1994}, spike formation on the droplet’s poles \cite{timonen_switchable_2013} and others.
It should be noted that droplets under the action of electromagnetic field have many uses such as  \cite{serwane_vivo_2017}, dynamic self assembly \cite{timonen_switchable_2013} and others.

In parallel with the experimental investigation of magnetic droplets, significant efforts in the development of the numerical methods for their simulation have been undertaken.
Efficient tools for the simulation of the  free boundary phenomena may be developed on the basis of the boundary integral equations \cite{pozrikidis-practical-2002, pozrikidis_boundary_1992}.
In axisymmetric case these methods were developed in \cite{sherwood_breakup_1988, stone_drops_1999}.
Among the phenomena predicted is, for example, the formation of the spikes on the droplet’s poles if the magnetic permeability is high enough \cite{stone_drops_1999, lavrova_numerical_2006}.
It may be noted that by using boundary integral equation technique the simulation of such complicated free boundary problem as the formation of the labyrinthine patterns in the Hele-Shaw cells has been carried out \cite{drikis99_labyrinth}.
The application of the boundary integral equation algorithm for the real three dimensional case is a real challenge since special care should be applied to keep the quality of mesh on the the droplets surface \cite{zinchenko_novel_1997, zinchenko_emulsion_2013, cristini_adaptive_2001}.
An example of the immersed boundary method application to modelling such dynamics is given by \cite{jesus2018deformation}.
A further review of magnetic fluid modelling and simulations is also available \cite{afkhami_ferrofluids_2017}.
In parallel to the development of the numerical tools for the simulation of magnetic droplets, corresponding elaborations are taking place for simulation of droplets in leaky dielectrics where besides the usual terms, the convective surface charge transfer by the liquid motion should be taken into account \cite{das_electrohydrodynamics_2017}.

One of the first undertakings to simulate magnetic droplets in the three dimensional case was undertaken in \cite{erdmanis_magnetic_2017} under the condition of equal viscosities of the droplet and surrounding fluid.
Since the viscosities of the concentrated phase of strongly magnetic droplets obtained by the demixing of magnetic colloids are significantly larger than the viscosity of the carrier liquid (usually water) \cite{stikuts_spontaneous_2020} it is crucial to account that in the numerical models.

At present there do not exist exact solutions of magnetic droplet behaviour under the simultaneous action of viscous, magnetic and capillary forces which may be used as benchmarks for validating numerical models.
In this situation the validations of the numerical models is carried out by the comparison of the numerical results with some approximate solutions.
It is our aim here to carry out these comparisons using the simple model of a magnetic fluid droplet  \cite{rosensweig_ferrohydrodynamics_1985} using an extension of the numerical algorithm of \cite{erdmanis_magnetic_2017}.

The paper is organized as follows. The mathematical model and the relevant equations are outlined in \S\ref{sec:math}. 
Next, section \S\ref{sec:algorithm} introduces the numerical algorithm, which is then validated with known theoretical relationships, as can be seen in \S \ref{sec:validation}. 
Finally, simulations of droplet behavior in different situations by the developed algorithm are presented in \S \ref{sec:results}, concluding with a short discussion.

\section{Mathematical model} 
\label{sec:math}
\subsection{Equations of motion}
We consider a droplet of magnetic fluid suspended in an infinite non-magnetic carrier fluid. We look at the case where the inertia of the fluid is negligible. Then the motion is governed by the Stokes equations for a magnetic fluid \cite{rosensweig_ferrohydrodynamics_1985, blums_magnetic_1997}.

\begin{equation}
    -\vect{\nabla}p + \eta \Delta \vect{v} + \vect{f_M}=\vect{0},\qquad \vect{\nabla}\cdot \vect{v}=0,
\end{equation}
where $p$ is the pressure, $\eta$ is the dynamic viscosity, ${f_M}_i=\partial_k T_{ik}$ is the volume force due to magnetic field, and $T_{ik}=-\frac{1}{2}\mu_0 H^2 \delta_{ik}+H_i B_k$ is the Maxwell stress tensor.
The boundary conditions for forces on the droplet surface read
\begin{equation}
    (\sigma_{ik}^{(e)}- \sigma_{ik}^{(i)})n_k + (T_{ik}^{(e)}-T_{ik}^{(i)})n_k - \gamma (k_1+k_2) n_i =0,
\end{equation}
where $\sigma_{ik} = -p\delta_{ik} + \eta\left(\partial_i v_k + \partial_k v_i\right)$ is the stress tensor of the fluid, $\gamma(k_1+k_2)$ is the capillary force due to the surface tension, and $\vect{n}$ is the unit normal vector pointing out of the droplet.
The superscripts $(e)$ and $(i)$ denote the parameters outside and inside of the droplet, respectively.

\subsection{Equations of motion in integral form}
It is possible to write an integral equation for the velocity of the points on the droplet's surface, which automatically satisfies the boundary conditions \citep{pozrikidis_boundary_1992}.

\begin{equation}
\begin{split}
	v_k(\vect{y}) =& \frac{1}{4\pi(\eta^{(e)}+\eta^{(i)})}\int_S f(\vect{x})n_i(\vect{x})G_{ik}(\vect{x},\vect{y})dS_x \\ &+ \frac{1}{4\pi}\frac{\eta^{(e)}-\eta^{(i)}}{\eta^{(e)}+\eta^{(i)}}\int_S v_i(\vect{x})T_{ijk}(\vect{x},\vect{y})n_j(\vect{x})dS_x + \frac{2\eta^{(e)}}{\eta^{(e)}+\eta^{(i)}}{v_0}_k(\vect{y}),
\end{split}
\end{equation}
where the integral is over the surface of the droplet, the integral kernels are the stokeslet $G_{ik}(\vect{x},\vect{y}) = \frac{\delta_{ij}}{|\vect{x}-\vect{y}|} + \frac{(x_i-y_i)(x_j-y_j)}{|\vect{x}-\vect{y}|^3}$, and the stresslet $T_{ijk}(\vect{x},\vect{y}) =  -6\frac{(x_i-y_i)(x_j-y_j)(x_k-y_k)}{|\vect{x}-\vect{y}|^5}$, $f(\vect{x})$ is the combined normal force of surface tension and magnetic forces on the droplet's surface, and $\vect{v}_0(\vect{y})$ is the ambient flow to which the droplet is subjected. 
The normal force on the surface can be written as \citep{erdmanis_magnetic_2017}
\begin{equation}
   f= \left( \frac{1}{2}\mu_0\mu(\mu-1){H^{(i)}}_n^2 + \frac{1}{2} \mu_0 (\mu-1) {H^{(i)}}_t^2 - \gamma (k_1+k_2) \right),
\end{equation}
where $H_t$ and $H_n$ are the tangential and normal components of the magnetic field on the surface, and $\mu$ is the relative permeability of the droplet.

\subsection{Dimensionless variables}
We introduce the length scale as the radius of a spherical droplet $R_0$, time scale $t_0=R_0\eta^{(e)}/\gamma$, a magnetic field scale $H_0$, a magnetic permeability scale $\mu_0$, a viscosity parameter $\lambda=\eta^{(i)}/\eta^{(e)}$ and the Bond magnetic number $Bm=4\pi\mu_0 R_0H_0^2/\gamma$.

Then the integral equations can be rewritten in a dimensionless form.

\begin{equation}
\begin{split}
	v_k(\vect{y}) =& -\frac{1}{1+\lambda}\frac{1}{4\pi}\int_S (k_1(\vect{x})+k_2(\vect{x}))n_i(\vect{x})G_{ik}(\vect{x},\vect{y})dS_x \\ 
	& +\frac{1}{1+\lambda}\frac{1}{4\pi}\int_S f_M(\vect{x}) n_i(\vect{x})G_{ik}(\vect{x},\vect{y})dS_x\\
	&+ \frac{1-\lambda}{1+\lambda}\frac{1}{4\pi}\int_S v_i(\vect{x})T_{ijk}(\vect{x},\vect{y})n_j(\vect{x})dS_x \\
	&+ \frac{2}{1+\lambda}{v_0}_k(\vect{y})
	\label{eq:v_inteq}
\end{split}
\end{equation}


The magnetic part of the normal force was separated out and in a dimensionless form reads 
\begin{equation}
f_M= Bm\left( \frac{1}{2}\mu(\mu-1){H^{(i)}}_n^2 + \frac{1}{2}(\mu-1) {H^{(i)}}_t^2\right).
\label{eq:surface_forces}
\end{equation}



\subsection{Regularization of the boundary integral equations}
\label{sseq:regularization}
All of the integrands in \eqref{eq:v_inteq} are weakly singular, they scale as $O(1/r)$ as $\vect x \rightarrow \vect y$ and are convergent ($\vect{r}=\vect{x}-\vect{y}$).
The asymptotic behavior can easily be seen by writing out the integrands in Taylor series for small $\vect r$ in local coordinates centered at $\vect y$.
It may be noted that $\vect{r}\cdot \vect n = O(r^2)$ as $\vect x \rightarrow \vect y$.
The integrals can be tackled numerically, for example, by calculating them on the singular elements in polar coordinates, where a factor of $r$ arises from the differential area element \cite{pozrikidis-practical-2002}. 
Nonetheless, easier implementation and greater precision of the numerical scheme can be achieved by removing these singularities, which for these integrals is possible.

Since the first integral in \eqref{eq:v_inteq} is over a closed surface, it can be transformed in the following form \citep{Zinchenko1999}
\begin{equation}
\begin{split}
\int_S(k_1(\vect{x}) + &k_2(\vect{x}))n_i(\vect{x}) G_{ik}(\vect{x},\vect{y})dS_x =\\ &
-\int_S \left(    r_i n_i(\vect{x})n_k(\vect{y})  +  r_i n_i(\vect{y})n_k(\vect{x})  +  (1 - n_i(\vect{x}) n_i(\vect{y}) )r_k \vphantom{\frac{(num)}{(\vect{d}en)}} \right. \\
 &- \left. \frac{3 r_k  (  n_i(\vect{x}) + n_i(\vect{y}) ) r_i r_j n_j(\vect{x}) }{|\vect{r}|^2}    \right)\frac{dS_x}{|\vect{r}|^3},
\end{split}.
\end{equation}
Note that the term in the brackets of the integral scales as $|\vect{r}|^3$ and the whole integrand scales as $O(1)$, as $\vect{x}\rightarrow \vect{y}$. 
Furthermore, calculating the sum of principal curvatures $(k_1+k_2)=\nabl \cdot \vect{n}$ in the first term on the right hand side of \eqref{eq:v_inteq} notoriously introduces large errors from the surface discretization \cite{zinchenko_novel_1997, Zinchenko1999}, which in this form is not necessary.

The singularity in the second integral in \eqref{eq:v_inteq} can be reduced by an order using singularity subtraction \cite{pozrikidis_boundary_1992}.
We use the identity \cite{erdmanis_magnetic_2017} $\int_S n_i(\vect{x}) G_{ik}(\vect x, \vect{y}) dS_x = 0$, which we multiply by $f_M(\vect y)$ and subtract from the second integral in \eqref{eq:v_inteq} to get
\begin{equation}
\begin{split}
    \int_S f_M(\vect{x})  n_i(\vect{x})G_{ik}(\vect{x},\vect{y})dS_x = \int_S [f_M(\vect{x}) - f_M(\vect{y}) ] n_i(\vect{x})G_{ik}(\vect{x},\vect{y})dS_x,
\end{split}
\end{equation}
where the integrand on the right hand side is $O(1)$ as $\vect x \rightarrow \vect y$, if $f_M(\vect{x})$ is smooth.

The third integral in \eqref{eq:v_inteq} is also treated by singularity subtraction using the identity \cite{pozrikidis_boundary_1992} $\int_S T_{ijk}(\vect x, \vect{y}) n_j(\vect{x}) dS_x = -4\pi\delta_{ik}$  and is replaced by
\begin{equation}
    \int_S v_i(\vect{x})T_{ijk}(\vect{x},\vect{y})n_j(\vect{x})dS_x = \int_S \left[v_i(\vect{x}) - v_i(\vect{y})\right]T_{ijk}(\vect{x},\vect{y})n_j(\vect{x})dS_x - 4\pi v_k(\vect{y}).
\end{equation}
As a result the singularity has been reduced by one order, and the integrand stays bounded as $\vect{x}\rightarrow \vect{y}$, if $v_i(\vect{x})$ is smooth.

We can write the integral equation \eqref{eq:v_inteq} for $\vect{v}$ in the form
\begin{equation}
v_k(\vect{y}) = \frac{\kappa}{4\pi} \int_S v_i(\vect{x})T_{ijk}(\vect{x},\vect{y})n_j(\vect{x})dS_x + F_k(\vect{y}),
\end{equation}
where $\vect{F}$ is independent of $\vect{v}$ and $\kappa=\frac{1-\lambda}{1+\lambda}$. It is known \cite{kim_microhydrodynamics:_1991} that the homogeneous part of this integral equation has an eigensolutions with eigenvalues $\kappa=\pm1$. 
The $\kappa=1$ eigensolution is a uniform expansion of the droplet and the $\kappa=-1$ eigensolution is an arbitrary rigid body motion of it. Therefore, for cases where $\lambda \gg 1$ or $\lambda \ll 1$ the integral equation is poorly conditioned. 
Magnetic droplets obtained by demixing of magnetic colloids are observed to have large viscosities ($\lambda\approx100$ \citep{bacri_behavior_1994}). Hence we use Wielandt's deflation \cite{kim_microhydrodynamics:_1991}, a procedure where we formulate an equivalent integral equation that does not have the unwanted eigensolutions. 
The regularized integral equations and Wielandt's deflation is further explained in the \ref{app:Wielandt}.

\subsection{Magnetic part}

As we consider equations of magnetostatics, $\nabl \times \vect{H} = 0$, the magnetic field can be expressed as a gradient of magnetic scalar potential $\vect{H} = \nabl\psi$. Since $\nabl \cdot \vect{H} = 0$ by the virtue of Maxwell's equation for $\vect{B}$ and assumed uniform magnetization $\vect{M}$, the magnetic potential satisfies the Laplace equation $\Delta\psi = 0$. 

Taking into account the continuity of the scalar potential $\psi^{(i)} = \psi^{(e)}$ and of the normal field component $\mu \nabl\psi^{(i)} \cdot \vect{n} = \nabl \psi^{(e)} \cdot \vect{n}$ on the fluid interface $S$, Laplace's equation can be recast in an integral equation form \citep{pozrikidis_boundary_1992}:
\begin{equation}
    \psi(\vect{y}) = \frac{2\vect{H_0}\cdot\vect{y}}{\mu+1} - \frac{1}{2\pi} \frac{\mu-1}{\mu+1} \int_{S} \psi(\vect{x}) \nabl_x\left(\frac{1}{r}\right)\cdot \vect{n}(\vect{x})\ \dd S_x,
    \label{eq:laplace_integral}
\end{equation}


where the $\vect{H_0}$ term represents an unperturbed background field.

The magnetic potential is calculated on discrete points describing the droplet, this allows us to calculate the tangential field component at each node $\vect{H_t}=|(\matr{I}-\vect{n} \otimes \vect{n})\ \nabla \psi|$.

Since the nodes are on the surface of the droplet, it is not possible to directly calculate the normal component of the gradient of the magnetic potential. 
Therefore, to calculate the normal component of the fild, we use the relation, where the normal field is expressed entirely in terms of the tangential field component \citep{erdmanis_magnetic_2017}:

\begin{equation}
    \vecf{H_n}{y} = \frac{\vect{H_0}\cdot\vecf{n}{y}}{\mu} - \frac{\mu-1}{4\pi\mu}\ \vecf{n}{y} \cdot \mathcal{P} \int_{S}  \left[ \big(\vecf{n}{x} \times \vecf{H_t}{x} \big) \times \nabl_x\left(\frac{1}{r}\right) \right]\ \dd S_x.
    \label{eq:Hn_integral}
\end{equation}
The integrand in \eqref{eq:Hn_integral} has a $O(1/r^2)$ singularity when $\vect{y} \rightarrow \vect{x}$, therefore it should be evaluated as a Cauchy principal value integral denoted by $\mathcal{P}$.
Similarly as before with velocity integral equations singularity subtraction is applied, making use of identities
\begin{align}
    &\int_S \frac{1}{4\pi} \nabl_x\left(\frac{1}{r}\right) \cdot \vect{n}(\vect{x})\ \dd S_x = -\frac{1}{2}, \label{eq:regular_id_1} \\
    &\int_S \frac{\vecf{n}{x} \times \vect{r}}{r^3}\   \dd S_x = \vect{0}, \label{eq:regular_id_2}
\end{align} 
where $\vect{y}$ lies on the boundary of the region of integration.

The equation for magnetic potential can be cast in a regularized form by means of \eqref{eq:regular_id_1}:
\begin{equation}
    \psi(\vect{y}) = \vect{H_0}(\vect{y})\cdot\vect{y} - \frac{\mu-1}{4\pi}\int_{S} \left[\psi(\vect{x}) - \psi(\vect{y}) \right] \nabl_x\left(\frac{1}{r}\right)\cdot \vect{n}(\vect{x})\ \dd S_x.
    \label{eq:laplace_integral}
\end{equation}

Relation for the normal field component is regularized by using \eqref{eq:regular_id_1} and \eqref{eq:regular_id_2} \citep{erdmanis_magnetic_2017} (here the sign is opposite in front of the integral term as compared to that in Eq. (3.9) of \cite{erdmanis_magnetic_2017}):

\begin{align}
    H_n(\vect{y}) = \frac{\vect{H_0} \cdot \vecf{n}{y}}{\mu} - \frac{\mu-1}{4\pi\mu}\ \vecf{n}{y} \cdot \int_{S} \Bigg[ \left[\vecf{H_t}{x} - \vecf{H_t}{y}\right]\  \nabl_x\left(\frac{1}{r}\right)\cdot \vect{n}(\vect{x}) - \nonumber \\ \nonumber       \left[\vecf{H_t}{x} - \vecf{H_t}{y}\right] \cross \left( \vecf{n}{x} \cross \nabl_x \left( \frac{1}{r}\right)  \right) \Bigg]\dd S_x.
\end{align}
Similarly as in the case of the velocity calculation, the integrands in the regularized equations are now bounded for $\vect{y}\rightarrow\vect{x}$, and can easily be integrated numerically. 
Singular integrands are calculated using local polar coordinates centered at $\vect{y}$ for the singular elements. \cite{pozrikidis-practical-2002, erdmanis_magnetic_2017}.

We also derived an alternative approach for calculating the normal and tangential field components from normal field differences on the surface, but it was observed to be less precise and slower and consequently was not used in further calculations.
For details, refer to \ref{app:normal_field_geometric} and \ref{app:tangential_field}.

Having solved these boundary integral equations for the magnetic field, we can obtain the forces acting of the droplet surface \eqref{eq:surface_forces} and subsequently calculate the velocity of each mesh node.

\section{Numerical algorithm} 
\label{sec:algorithm}

The numerical approach  is an extension of \cite{erdmanis_magnetic_2017}, where only equilibrium droplet configurations were calculated at equal fluid viscosities.

The magnetic fluid droplet surface is triangulated by a mesh of boundary elements with collocation points or nodes.
The integrals are solved using the trapezoid integration scheme using only the function values on the nodes. 
This allows us to conveniently reformulate the summation over the flat triangles to the summation over nodes \citep{zinchenko_novel_1997}
\begin{equation}
    \int_S f(\vect x)dS \approx \sum_i f(\vect x_i) \Delta S_i,
\end{equation}
where the summation is done over all nodes $i$, and $\Delta S_i=1/3 \sum \Delta S$ is the average area of the three triangles adjacent to the node $i$. 
This way the integral equations become linear systems of equations that can be solved using common numerical libraries.
Solution of the integral equations of \S \ref{sec:math} gives the velocities of nodes and allows us to calculate the dynamics of droplet shapes.

The original spherical mesh is generated by iteratively ``growing'' an icosahedron, by adding more nodes in its faces and projecting the new nodes on a sphere, as proposed in \citep{siqueira_new_2017}.
The normals and curvatures on each vertex are found by fitting a paraboloid on the relevant vertex and its immediate neighbors \citep{zinchenko_novel_1997}.

\subsection{Mesh maintenance}

Multiple mesh stabilization techniques are employed throughout the simulations, as the mesh tends to degrade rather quickly. The methods utilized in this paper are explained below.

\subsubsection{Passive stabilization}
Since the dynamics of the droplet shape is determined by the normal velocity component only, the mesh may be stabilized by using proper tangential velocity components.
In \textit{passive stabilization} \cite{zinchenko_emulsion_2013}, the tangential components can be adjusted in order to minimize a ``kinetic energy" function
\begin{equation}
    F = \sum_{\vect{x}_{ij}} \left[ \frac{d}{dt} \left( \frac{|\vect{x}_{ij}|^2}{h^2_{ij}} + \frac{h^2_{ij}}{|\vect{x}_{ij}|^2}\right)\right]^2 + 0.4 \sum_\Delta \frac{1}{C^2_\Delta} \left( \frac{dC_\Delta}{dt} \right)^2,
\end{equation}
where the first sum pertains to edges between nodes $i$ and $j$ and tries to keep edges at the optimal lengths of $h_{ij}$ as determined by local curvatures \cite{zinchenko_emulsion_2013}, while the second sum pertains to mesh surface triangles and tries to keep the triangles as regular as possible, using the ``compactness" of a triangle $C_\Delta = S_\Delta/(a^2+b^2+c^2)$, with $a,b,c$ representing the lengths of its sides, as a guide \cite{zinchenko_emulsion_2013}.

\subsubsection{Active stabilization}
The above algorithm slows the mesh degradation but does not stop it completely. In addition \textit{active stabilization} \cite{zinchenko_emulsion_2013} between iterations is necessary whereby the nodes are translated along the surface of the droplet in order to minimize a ``potential energy" function
\begin{equation}
    E = \sum_{\vect{x}_{ij}} \left[ \frac{1}{2} \left( \frac{|\vect{x}_{ij}|^2}{h^2_{ij}} + \frac{h^2_{ij}}{|\vect{x}_{ij}|^2}\right)\right]^{50} + \sum_\Delta \left( \frac{C^{reg}_\Delta}{C_\Delta} \right)^{100},
\end{equation}
where $C^{reg}_\Delta = \sqrt{3}/12$ is the compactness value of a regular triangle. 
This function $E$ assumes large values when $\vect{x}_{ij}$ differs a lot from its optimal length of $h_{ij}$ and so it avoids both unwarranted crowding and dispersion of vertices, as well disallowing triangles to deviate a lot from the optimal compactness of a regular triangle. 
In order to translate the node $\vect{x}^i$ along the droplet surface, we use the previously fitted paraboloid used in the normal vector calculations  with $\vect{x}^i$ lying at its tip, locally approximately coinciding with the droplet surface, as described by its neighboring vertices. The node $\vect{x}^i$ is then translated along this surface in order to minimize the ``potential energy" function $E$.

The above energy functions $E$ and $F$ can have their gradients expressed in an explicit analytical form, allowing for efficient optimization. 
We use the conjugate gradient method to find the minimum of $F$. And to minimize $E$, we use a modified gradient descent, where after each step we project the points back on the paraboloid that describes the local droplet surface \cite{zinchenko_emulsion_2013}.

\subsubsection{Edge flipping}
Edges between vertices may be flipped. Consider a quadrilateral described by vertices $\left[ \vect{x}^a, \vect{x}^b, \vect{x}^d, \vect{x}^c \right]$ with an additional edge, connecting the two diagonally opposite vertices $\left[\vect{x}^a, \vect{x}^d \right]$. This edge might be flipped to instead connect the vertices $\left[\vect{x}^b, \vect{x}^c \right]$, provided that
\begin{equation*}
    \left|\vect{x}^b - \vect{x}^c \right|^2 < \left|(\vect{x}^b-\vect{O}^b)(\vect{x}^c - \vect{x}^b)\right| + 
    \Big|(\vect{x}^c-\vect{O}^c)(\vect{x}^c - \vect{x}^b)\Big|,
\end{equation*}
with $\vect{O}^k$ being the circumcenter of the triangle $\left[ \vect{x}^a, \vect{x}^d, \vect{x}^k \right]$, where $k \in \{b,c\}$
\cite{zinchenko_emulsion_2013,cristini_adaptive_2001}. Such flips allow for increasingly regular triangles and they have to be applied iteratively to all of the edges until no more flips are possible. 
An edge will not be flipped, if it would result in a node with less than 5 connected nodes. 
This is done to ensure that a general paraboloid can be fitted on every node and its neighbors.
Edge flipping is applied at each simulation step. If an edge is flipped, active stabilization is applied again.

\subsubsection{Node addition}
We found that to achieve accurate results, a larger number of nodes are needed in regions of high curvature than can be sustained with the above mentioned techniques, therefore, we employ a node addition routine. 
For every vertex, we compute the magnitude of curvature $H=\sqrt{k_1^2+k_2^2}$, where $k_1, k_2$ are the principal curvatures we obtained from the fitted paraboloids. 
Then for each triangle we compute the mean curvature of its vertices $H_\Delta = \frac{1}{3} \sum\limits_{i \in \Delta} H_i$ and the square root of its area $\sqrt{S_\Delta}$, signifying the length scale of the triangle. 
Each triangle with $H_\Delta \sqrt{S_\Delta} > \varepsilon$ is marked for splitting, with $\varepsilon$ being an empirical cut-off criteria. 
Furthermore, if a triangle has two or more marked neighboring triangles, it also is marked.

\begin{figure}
    \centering
    \begin{subfigure}[c]{0.49\textwidth}
        \begin{overpic}[width=\textwidth]{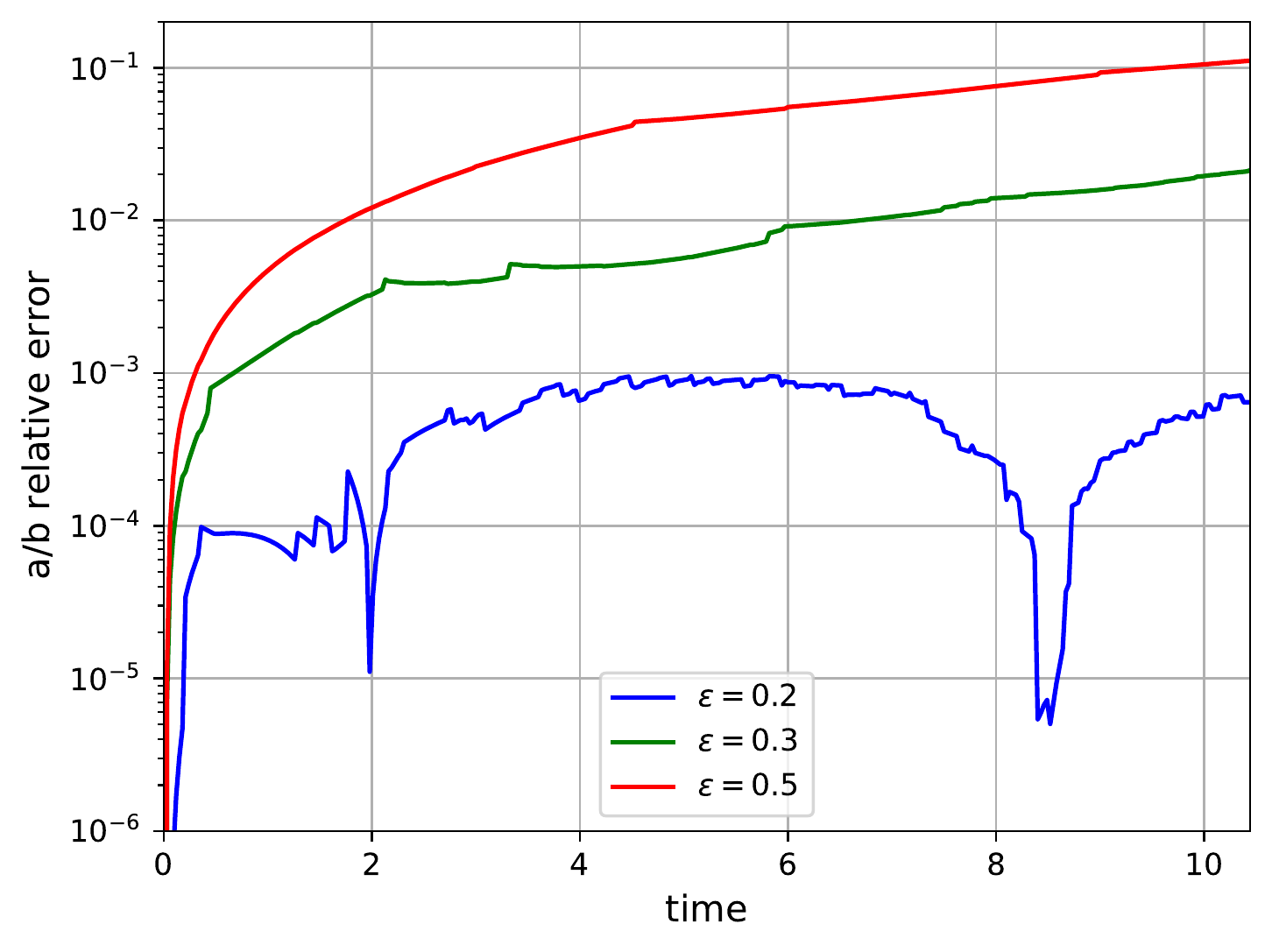}
            \put (-2,70) {\textcolor{black}{a)}}
        \end{overpic}
    \end{subfigure}
    \begin{subfigure}[c]{0.49\textwidth}
        \begin{overpic}[width=\textwidth]{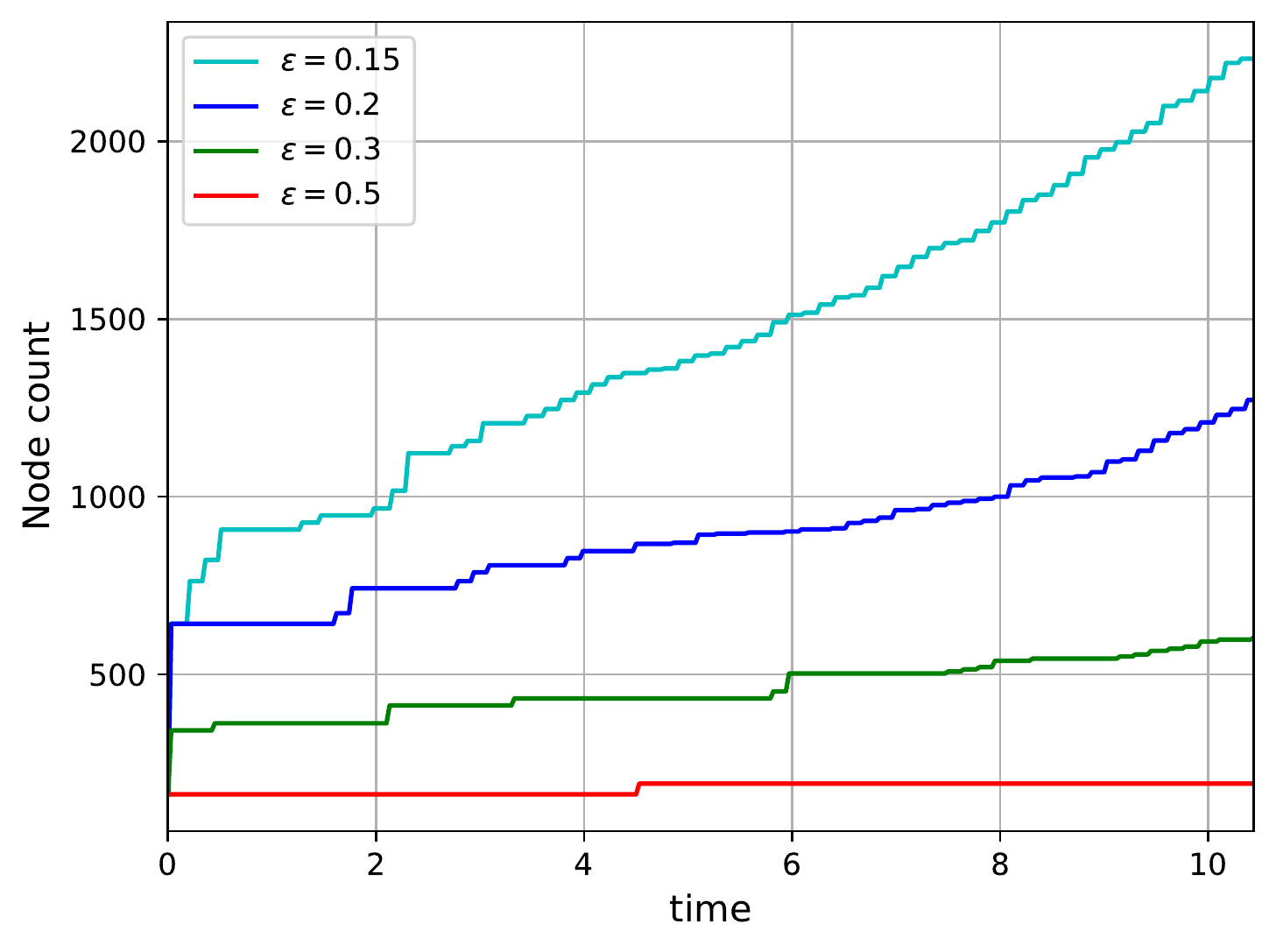}
            \put (0,70) {\textcolor{black}{b)}}
        \end{overpic}
    \end{subfigure}
    \caption{An initially spherical droplet with $\lambda=1$, $\mu=22$ is being elongated in a constant magnetic field such that $Bm=5.8$. The simulation time step is sufficiently small $\Delta t=0.03$. a) depicts the relative error of the droplet axis ratio with different cut-offs for node addition $\varepsilon$ compared to $\varepsilon=0.15$. The use of $\varepsilon=0.2$ was chosen for further simulations unless stated otherwise, it introduces an error of around $0.1\%$. b) depicts the number of nodes as the simulation progresses for different values of $\varepsilon$.}
    \label{fig:split_criterion}
\end{figure}

To determine an optimal $\varepsilon$ value, a droplet was stretched in a constant field with different $\varepsilon$ values and the ratio of the droplet semi-axes evolution compared (Figure \ref{fig:split_criterion}).
A threshold for cut-off of 0.2 was determined to be sufficient for a precision of $0.1\%$.
Smaller values of $\varepsilon$ quickly increase the number of nodes and thus significantly increase the computation time, which scales as roughly as $O(N^2)$, where $N$ is the number of nodes.

\begin{figure}
    \centering
    \begin{subfigure}[c]{0.45\textwidth}
        \begin{overpic}[width=\textwidth]{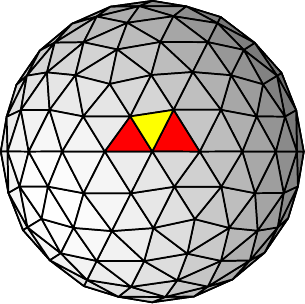}
            \put (20,105) {\textcolor{black}{before node addition}}
        \end{overpic}
    \end{subfigure}
    \hspace{2mm}
    \begin{subfigure}[c]{0.45\textwidth}
        \begin{overpic}[width=\textwidth]{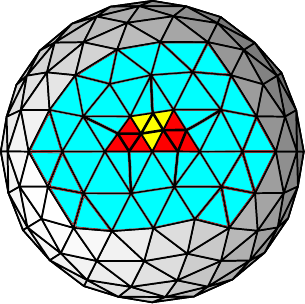}
            \put (20,105) {\textcolor{black}{after node addition}}
        \end{overpic}
    \end{subfigure}
    \caption{Before node addition the red triangles are marked for splitting. Since the yellow triangle has two neighbors that are marked for splitting, it also gets marked. After node addition, new triangles are created in such a way to ensure that there are at least 5 neighbors for each node. The new triangles and those (marked in blue) that are ``two layers" around the triangles marked for splitting are actively stabilized to improve the resultant mesh.}
    \label{fig:split_demonstration}
\end{figure}

The marked triangles will have new nodes added at the midpoint of each of their edges \cite{cristini_adaptive_2001}. 
The added nodes are mutually connected in such a way that all nodes have at least five neighbors.
Each of the new nodes will also be projected on the one of the original triangle's node's paraboloid to which  the new node is the closest to. 
The neighborhood of triangles affected by this addition is also actively stabilized similarly to \cite{cristini_adaptive_2001}.
The node addition procedure is illustrated in Figure \ref{fig:split_demonstration}.

\subsubsection{Time integration scheme}
We found space--discretization has a larger impact than time--discretization on convergence.

Therefore a simple Euler method for time integration is used, however it introduces some volume change in each iteration.
We remedy this by rescaling the volume at every iteration.
We use an adaptive time step $\Delta t$ inspired by \cite{zinchenko_algorithm_2008} 
\begin{equation}
\begin{aligned}
    \Delta \tau = 7.4\ \text{min}\left\{ \left(\frac{\Delta x_{min}}{|k|_{max}}\right)_i \right\}\\
    \Delta t = \text{min}\left\{ \Delta \tau, 0.05\frac{2\pi}{\omega}, 0.07 \right\}
\end{aligned},
\end{equation}
where $\Delta x_{min}$ is the shortest edge next to the node $i$, and $|k|_{max}$ is the largest principal curvature by absolute value at node $i$ and $\omega$ is the dimensionless magnetic field rotation frequency.

\subsubsection{Summary}
The numerical algorithm can be briefly summarized as follows:

\begin{itemize}
    \item For the given external magnetic field, solve the boundary integral equation for the magnetic potential $\psi$ on the droplet surface \eqref{eq:laplace_integral}.
    
    \item Calculate the tangential field component $\vect{H_t}=|(I-\vect{n} \otimes \vect{n})\ \nabla \psi|$. Express the normal field component $\vect{H_n}$ in a regularized form in terms of the tangential component \eqref{eq:Hn_integral}.

    \item Find the magnetic surface forces $\vect{f_M}$ \eqref{eq:surface_forces}.
    
    \item Nodes are moved by the first-order Euler algorithm according to the velocities found by solution of equation \eqref{eq:v_inteq} and adjusted by \textit{passive stabilization}. Afterwards, rescale droplet volume.
    
    \item Mesh maintenance via node addition and edge flipping are applied at every simulation step and \textit{active} stabilization is utilized every hundred iterations, unless any nodes have been added or any edges have been flipped, if so, it is used immediately.
\end{itemize}

\section{Algorithm validation} 
\label{sec:validation}

The algorithm has been validated with known theoretical relationships for droplet equilibrium configurations and dynamics when available.

\subsection{Relaxation to a sphere}
\begin{figure}
\centering
\begin{minipage}[t]{.47\textwidth}
    \centering
    \includegraphics[width=0.97\linewidth]{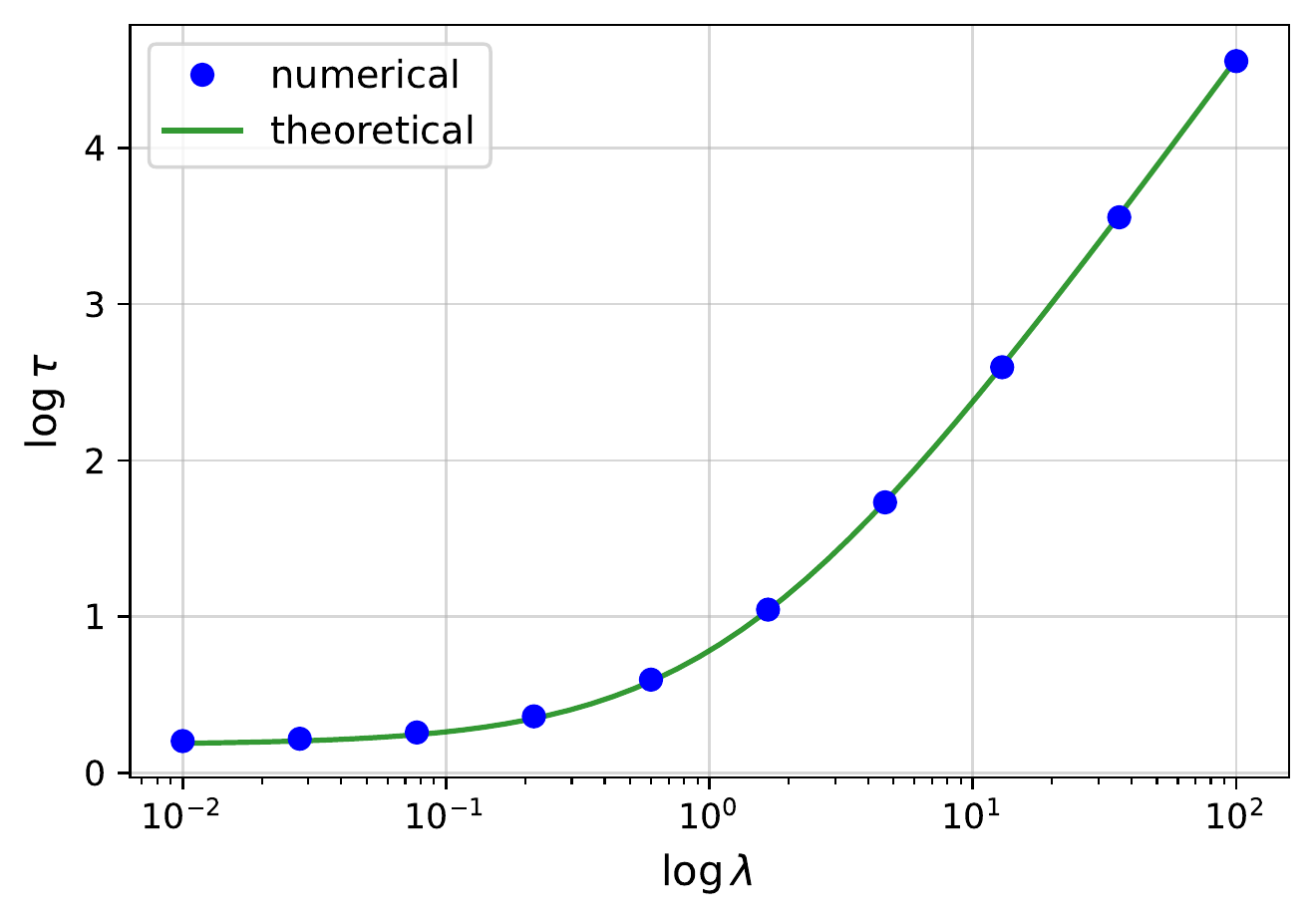}
    \caption{Characteristic relaxation dimensionless time $\tau$ of an elongated droplet depending on the droplet/fluid viscosity ratio $\lambda$. The points are the numerical results that closely follow the theoretical curve \eqref{eq:tau_relax}. }
    \label{fig:logtau_lambda}
\end{minipage}%
\hspace{1mm}
\begin{minipage}[t]{.5\textwidth}
    \centering
    \includegraphics[trim={0mm 0 0 9mm},clip,width=1.05\textwidth]{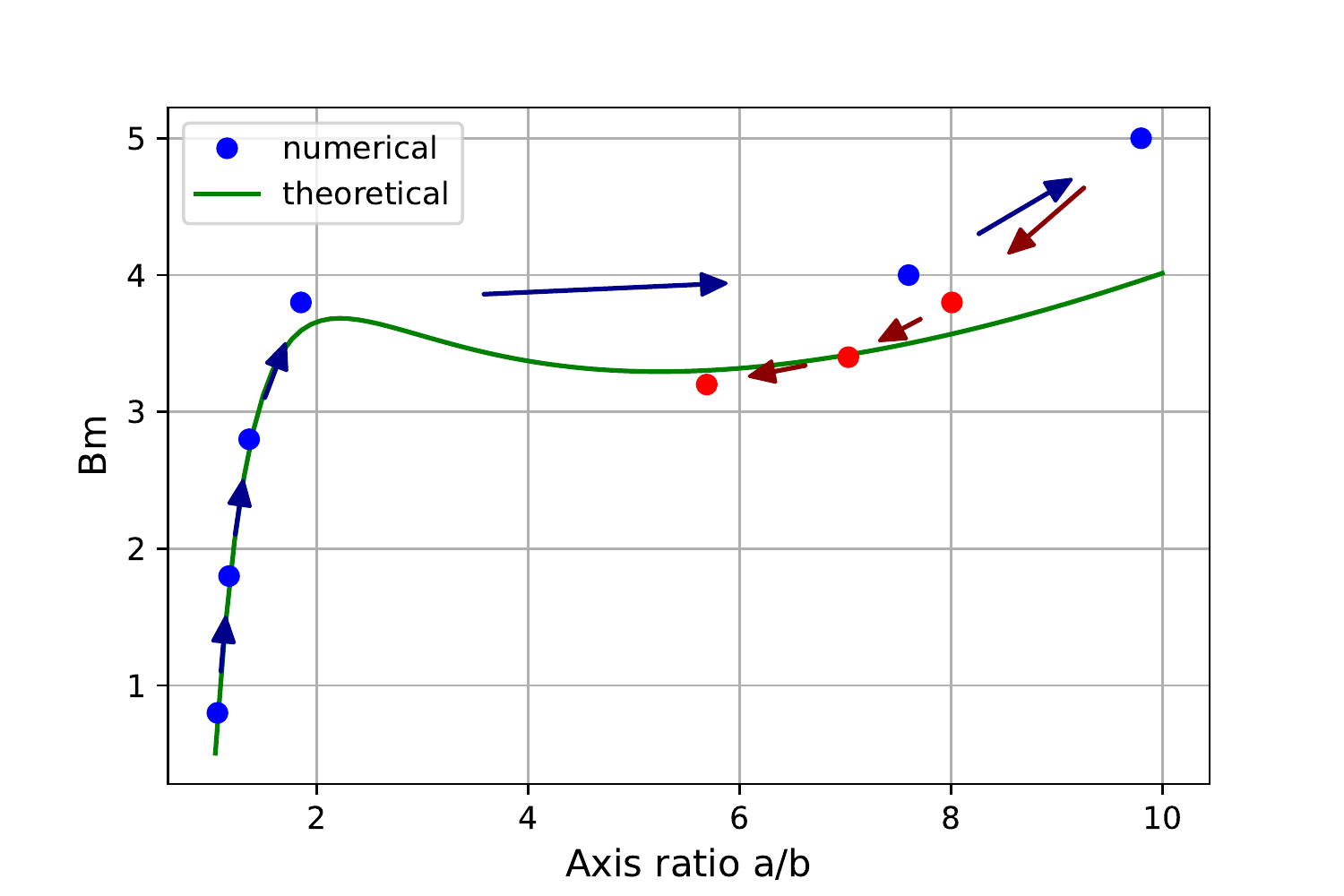}
    \caption{Evolution of droplet through the hysteresis region. The blue (red) points and arrows indicate the calculated hysteresis trajectory with increasing (decreasing) magnetic field once the droplet has equilibriated. The simulation used $\mu=30$ and $\lambda=7.6$.}
    \label{fig:hysteresis}
    \end{minipage}
\end{figure}

An elongated droplet approximated as an ellipsoid of rotation can be described by the Taylor deformation parameter parameter
\begin{equation}
    D = \frac{a-b}{a+b},
\end{equation}
where $a$ and $b$ are its major and minor semi-axis respectively. 
In the absence of an external field, small elongations decay exponentially due to capillary forces $D\propto e^{-t/\tau}$, where the characteristic relaxation time $\tau$, renormalized by the time scale of the system, reads \cite{DIKANSKY}:
\begin{equation}
\label{eq:tau_relax}
    \frac{\tau}{\eta^{(e)}R_0/\gamma}= \frac{(16+19\lambda)(3+2\lambda)}{40(1+\lambda)}.
\end{equation}
The numerically determined relaxation times are presented in Figure \ref{fig:logtau_lambda}.

\subsection{Equilibrium elongation in constant field}

We can further compare its equilibrium configurations against relations given below.
Figure \ref{fig:hysteresis} shows the comparison between the calculated equilibrium shapes of the droplets in a given magnetic field with the theoretical relation from \cite{bacri_instability_1982, cebers_virial_1985}
\begin{equation}
    Bm=\left[\frac{4\pi}{\mu-1}+N\right]^2\frac{1}{2\pi} 
    \frac{\left( \frac{3-2e^2}{e^2} - \frac{(3-4e^2)\arcsin{e}}{e^3(1-e^2)^{1/2}}\right)}
    {(1-e^2)^{2/3} \left( \frac{(3-e^2)}{e^5} \log{\left( \frac{1+e}{1-e}\right)} - \frac{6}{e^4}\right)},
    \label{eq:hysteresis_eq}
\end{equation}
where $e$ represents the eccentricity of the ellipsoid $e=\sqrt{1-b^2/a^2}$ with $b,\ a$ being its short and long semi-axis, respectively, and $N$ is the demagnetizing factor that for prolate ellipsoids reads
\begin{equation}
    N=\frac{4\pi(1-e^2)}{2e^3} \left( \log{\frac{1+e}{1-e}} - 2e \right).
\end{equation}

The equilibrium relation \eqref{eq:hysteresis_eq} is derived for an axisymmetric ellipsoidal droplet, an approximation which holds until the axial ratio of about 7 \cite{osti_6426307}.
This limit is also evidenced in Fig. \ref{fig:hysteresis} where the numerical result deviates at large droplet elongation, an effect explained by the droplet developing sharper tips than a fitted ellipsoid would have at the corresponding elongation, and thus no longer yielding to the ellipsoidal approximation.

As the simulation approaches equilibrium, the collocation points are moved by progressively smaller displacements.
To calculate the $t \rightarrow \infty$ behaviour, the points shown are obtained by using the Shanks transformation \citep{rallison_numerical_1981}.

There is a qualitative change in the equilibrium curve \eqref{eq:hysteresis_eq} that occurs with increasing permeability $\mu$ values. At $\mu \gtrsim 21$ the equilibrium curve becomes multivalued with respect to magnetic field, indicating an instability onset (Figure \ref{fig:hysteresis}).
If in this particular case with $\mu=30$, $Bm$ is increased past the critical value, $Bm_c\approx3.68$, the droplet configuration becomes unstable and has to ``jump'' to a new stable condition i.e. suddenly stretch. 
Once the droplet has reached this new stable configuration, the field can be lowered below the critical value, however the droplet will not ``jump'' back (contract) to its previously stable configuration, but rather slowly trace the equilibrium curve. This phenomenon is called hysteresis, whereby the system reverts to a state other than its original, when external perturbations have returned to their initial values.

With decreasing fields, the droplet would trace the curve until a second critical field value of around $Bm\approx3.32$ (for $\mu=30$) is reached. With even lower fields, the droplet would once again suddenly ``jump'' back (contract). 
Figure \ref{fig:hysteresis} shows the hysteresis path calculated by the algorithm.
In these simulations node addition was disabled since accurate description of such highly elongated droplets would require prohibitively many points, other mesh maintenance techniques were still employed.
Therefore the results should be interpreted only qualitatively.

\subsection{Elongation of quasi-stable droplets}
\subsubsection{Elongation bottleneck}

The closer the field is over the critical value, the longer the droplet will spend in this quasi-stable state, before ``jumping'' over to a truly stable configuration, indicating a time bottleneck region. The dynamics of this ``jump'' instability are governed by a hyperbolic equation shown by \cite{bacri_dynamics_1983} for small $t$
\begin{equation}
    \frac{a}{b}-\left(\frac{a}{b}\right)_c = S\ \tau \tan{\frac{t}{\tau}},
\label{eq:tange}
\end{equation}
where $a/b$ is semi-axial length ratio, here $\tau$ is the characteristic time spent in the bottleneck region before the ``jump'', $t$ represents time, $S$ is a numerical constant, and the subscript $c$ indicates critical value, i.e. the one at the extremum of the equilibrium curve \eqref{eq:hysteresis_eq}.

\begin{figure}[h]
\centering
\begin{minipage}[t]{.49\textwidth}
    \centering
    \includegraphics[trim={0mm 0 0 0mm}, clip,width=0.95\linewidth]{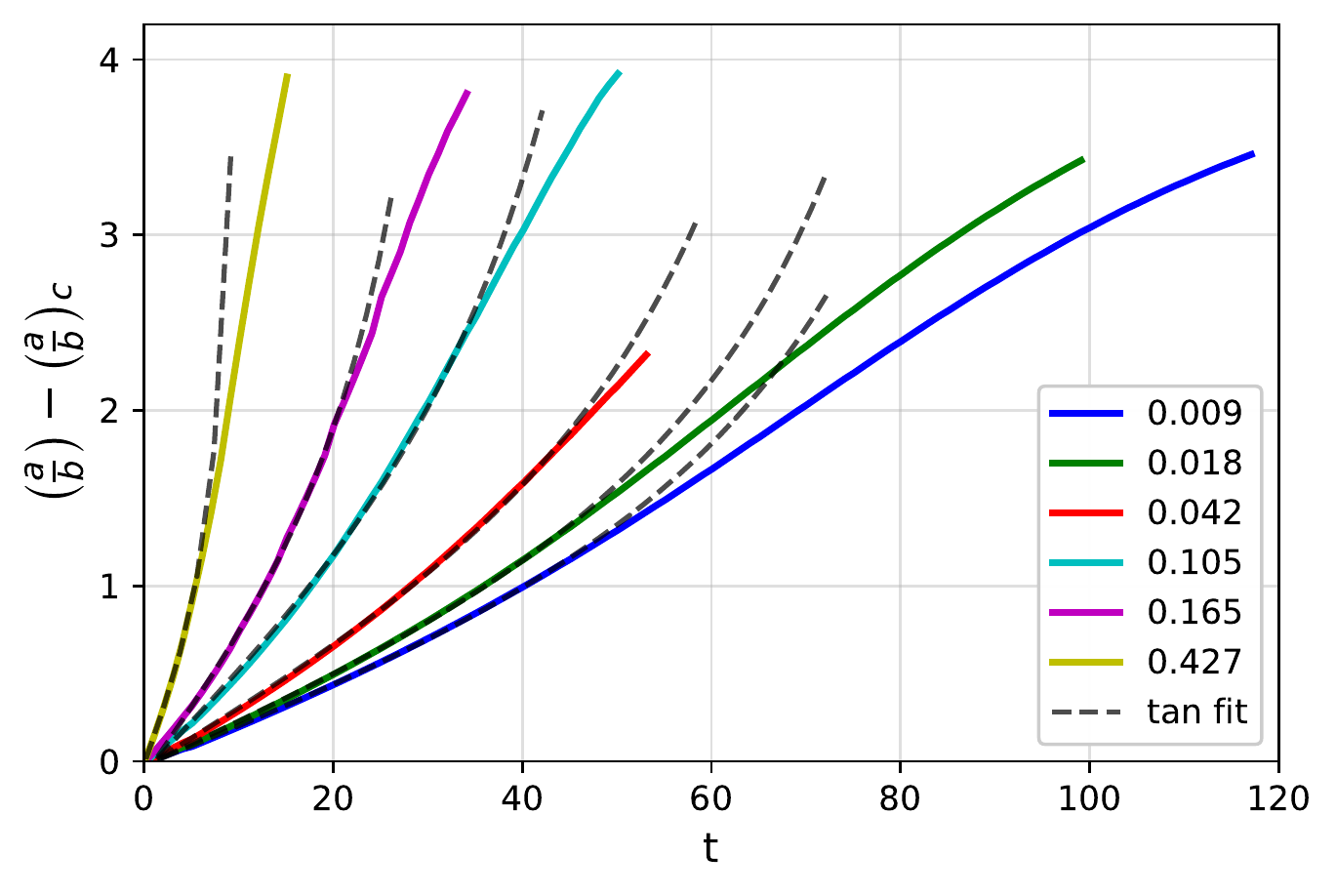}
    \caption{Elongations of the droplet whilst ``jumping over`` the hysteresis region at various external field $h$ values, with an overlaid tangential fits, according to \eqref{eq:tange}. The initially accelerated dynamics saturate as the droplet reaches its new equilibrium. Simulated using $\mu=30, \lambda=7.6$.}
    \label{fig:tange}
\end{minipage}%
\hspace{1mm}
\begin{minipage}[t]{.48\textwidth}
    \centering
    \includegraphics[trim={0mm 2mm 0 0mm},clip,width=1.0\linewidth]{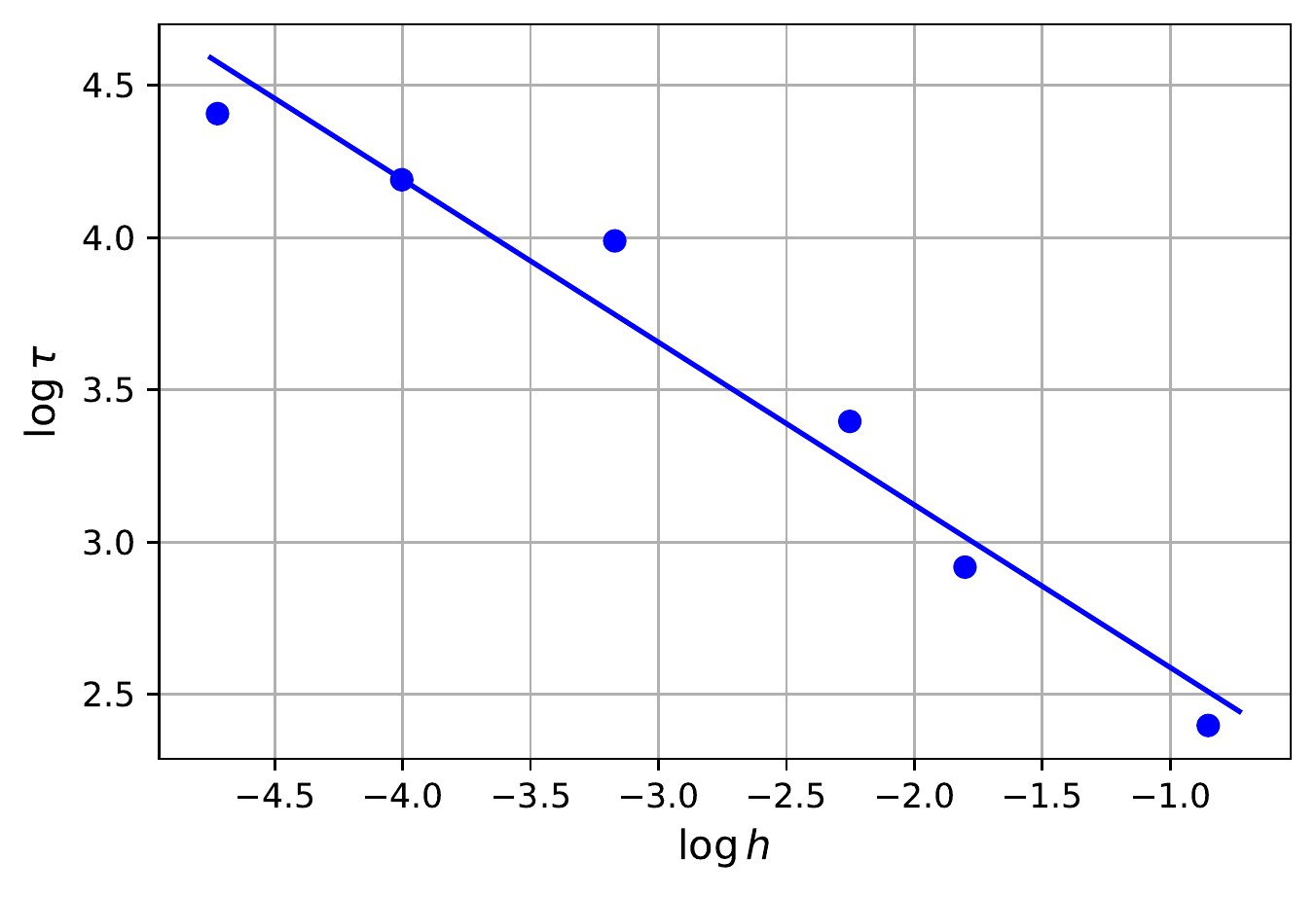}
    \caption{Characteristic time $\tau$ spent in the bottleneck region before a hysteresis jump with different magnetic fields that stretch the droplet. $h=\frac{H}{H_c}-1$. The points are obtained from numerical simulations and the line is the linear fit with a slope of $k=-0.534$, closely following the theoretical slope of $-0.5$ \citep{bacri_dynamics_1983}.}
    \label{fig:logt_logh}
\end{minipage}
\end{figure}

Examples of multiple droplet trajectories at different fields ${h=H/H_c - 1}$ are shown in Fig. \ref{fig:tange}, with the corresponding tangential fits \eqref{eq:tange} of $\tau$ and $S$ overlayed.
The trajectories were obtained by putting an ellipsoidal droplet ($\mu=30,\ \lambda=7.6$) elongated to the corresponding critical (pre-hysteresis ``jump'') axis ratio of $a/b=2.21$ and letting it evolve in various field strengths.
It can be observed that the initial slow dynamics indicating the ``bottleneck'' region are followed by a rather quick stretching -- the ``jump'' -- afterwards finally saturating into a new equilibrium position.

Figure \ref{fig:logt_logh} shows the numerically determined bottleneck behaviour of the droplet close to the critical field parameter $Bm_c$  before ``jumping" over the instable region to a stable configuration. 
The time spent in the ``bottleneck'' is expected to follow $\tau \sim \frac{1}{\sqrt{h}}$, where $h=\frac{H}{H_c}-1=\sqrt{\frac{Bm}{Bm_c}}-1$,  or $\log{\tau}\sim -0.5 \log{h}$ in logarithmic terms \cite{bacri_dynamics_1983}, which as shown in Fig. \ref{fig:logt_logh} is in good agreement with the value of $k=-0.534$ determined from numerical simulation.

\subsubsection{Virial theorem approach}

Analysis of this dynamics allows one to probe the applicability of  another theoretical magnetic fluid droplet description which is based on the Rayleigh dissipation function and the virial theorem. 
In the bottleneck region the sum of its surface and magnetic energies has stationary point of inflection with respect to its eccentricity $\partial_e E = \partial^2_{ee} E = 0$ \cite{bacri_instability_1982}. This allows for an approximation of bottleneck dynamics around the critical point in terms of unspecified constants \cite{bacri_dynamics_1983}:
\begin{equation}
    D\dot{e} = Ah + B(e-e_c)^2,
\end{equation}
with the constants taken at $e=e_c$ and $H=H_c$:
\begin{equation}
    D = -\eta R_0^3 f(e_c),\ \ \ \ A = H_c \frac{\partial^2 E}{\partial H \partial e},\ \ \ \ B = \frac{1}{2} \frac{\partial^3 E}{\partial e^3},
\end{equation}
with $\eta$ being the droplet viscosity and $f(e)$ an unknown function of the droplet shape, to be determined from an interplay of the Rayleigh dissipation function and the virial theorem.
Further, taking into account boundary conditions leads to the explicit expressions for the constants, as shown in \ref{app:virial}:
\begin{equation*}
    D = \frac{8 e_c}{9 (1-e_c^2)^{1/2}},\ \ \ 
    A =  \frac{8 \pi}{3}\ Bm_c\ \frac{6e_c + (e_c^2-3) \log \left[ \frac{1+e_c}{1-e_c}\right]}{2e_c^4}\ \ \ \ 
    B =  -\frac{1}{2} \frac{\partial^3 E}{\partial e^3} \frac{(1-e_c^2)^3}{e_c^2}.
\label{eq:ABD}    
\end{equation*}

Fitting \eqref{eq:tange} to numerical data of the ``bottleneck'' region dynamics gives the value of the constant ${S=A/(2D)(H^2/H^2_c-1)}$ \cite{bacri_dynamics_1983}, afterwards allowing to extract the $A/2D$ ratio from the $S$ vs. $H^2$ linear fit.
Furthermore, fitting ${\tau = \tau_0/\sqrt{h}}$ \cite{bacri_dynamics_1983} to the characteristic times spent in the ``bottleneck'' region $\tau$ (Figure \ref{fig:logt_logh}) gives the estimate for the capillary time ${\tau_0=\sqrt{AB}/D}$.

A comparison of determined calculated constant values is shown in Table \ref{tab:comparison}.
\begin{table}[h!]
\begin{center}
\begin{tabular}{|c|c c c|} 
 \hline
   & Bacri84 & Virial th. & Numerical \\ [0.5ex] 
 \hline
 $A\ /\ 2D $& 0.73 & 1.18 & 0.98 \\ 
 \hline
 $\sqrt{AB}\ /\ D$ & 0.9 & 0.6 & 1.0 \\
 \hline
 
\end{tabular}
\end{center}
\caption{Comparison of determined constant values with various approaches.}
\label{tab:comparison}
\end{table}

\section{Simulations}
\label{sec:results}

\subsection{Constant field}

\begin{figure}[h]
\centering
    \begin{subfigure}[c]{.24\textwidth}
        \begin{overpic}[trim={8cm 4cm 8cm 4cm},clip,width=\textwidth]{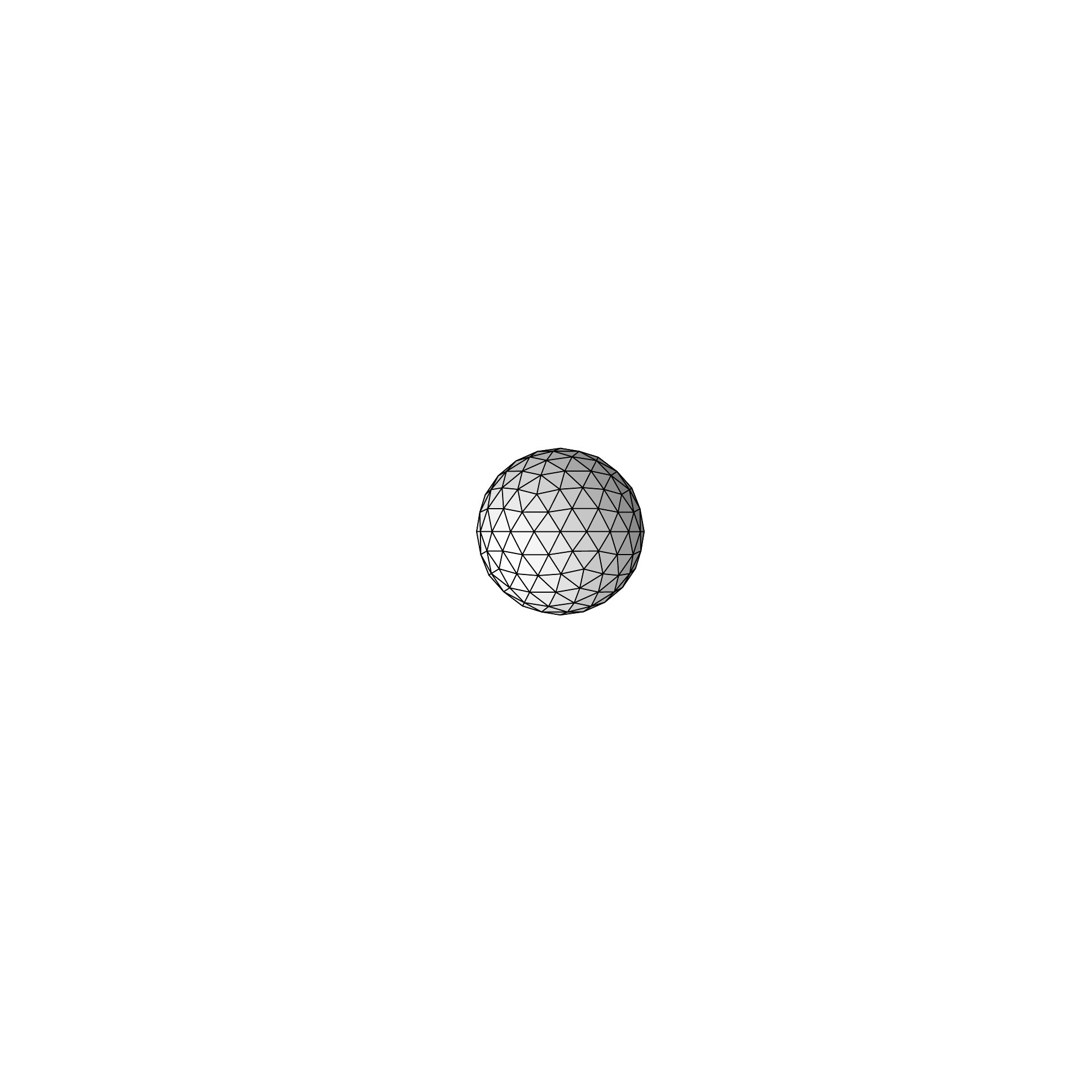}
            \put (0,70) {\textcolor{black}{$t=0.6$}}
        \end{overpic}
    \end{subfigure}
    \begin{subfigure}[c]{.24\textwidth}
        \begin{overpic}[trim={8cm 4cm 8cm 4cm},clip,width=\textwidth]{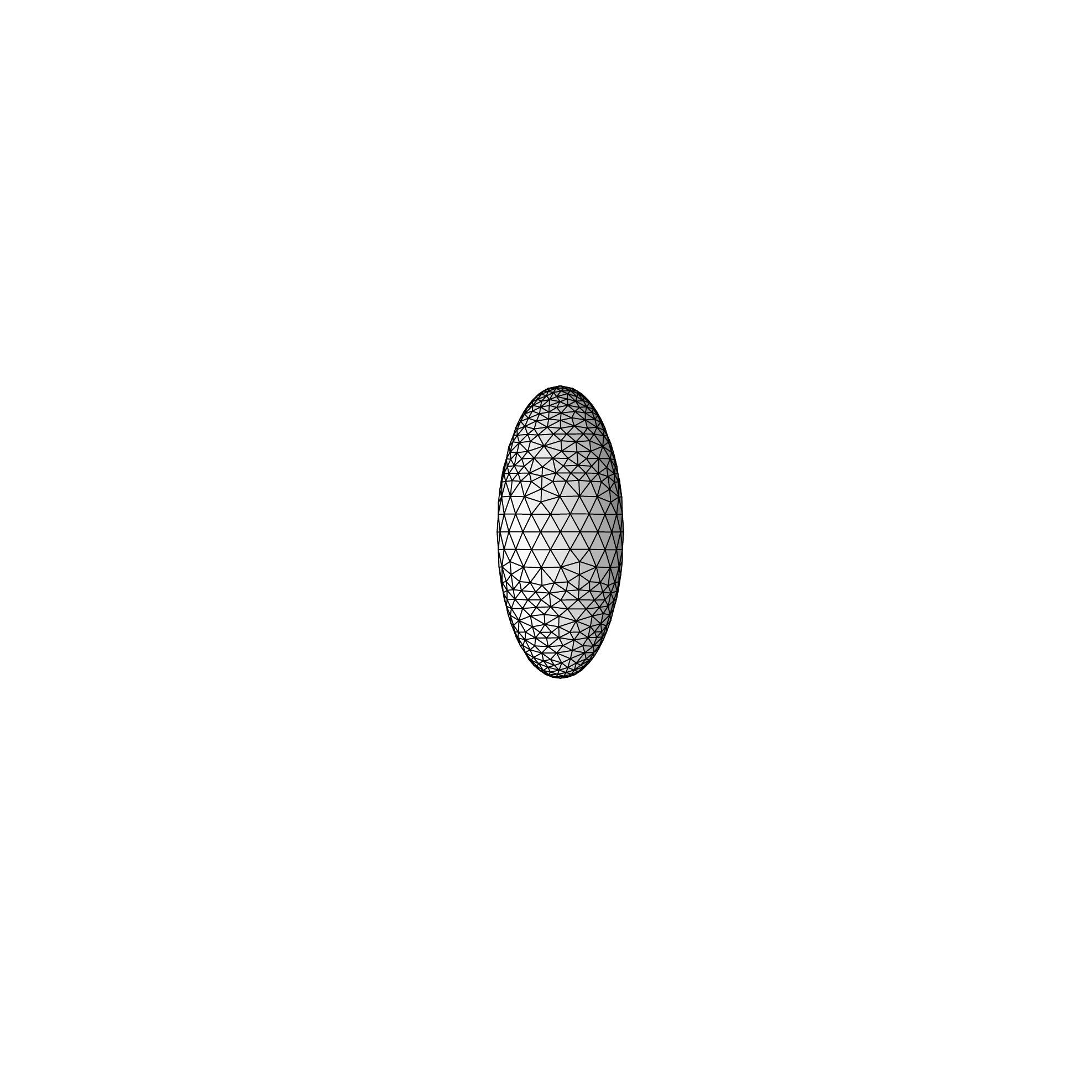}
            \put (0,70) {\textcolor{black}{$t=47$}}
        \end{overpic}
    \end{subfigure}
    \begin{subfigure}[c]{.24\textwidth}
        \begin{overpic}[trim={8cm 4cm 8cm 4cm},clip,width=\textwidth]{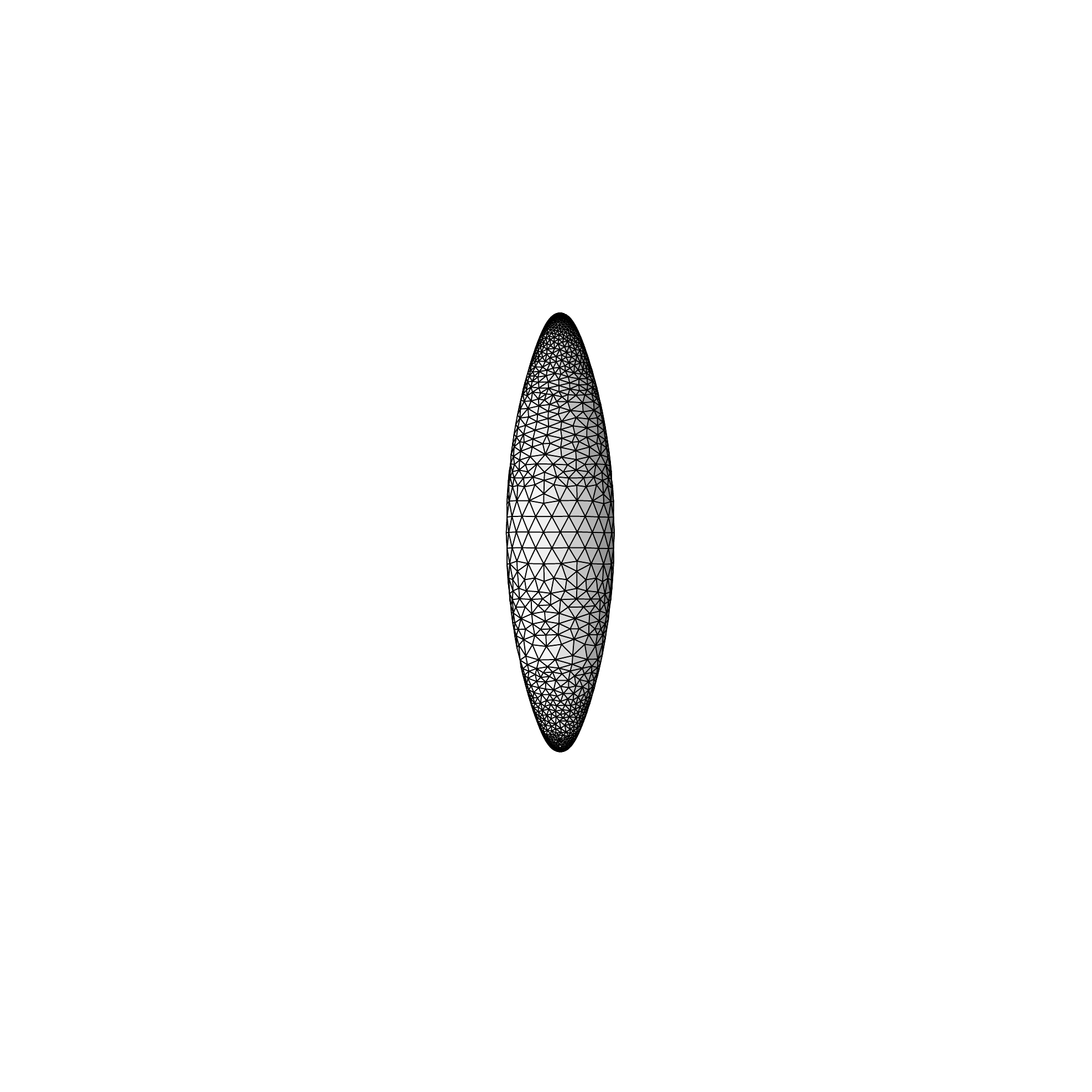}
            \put (0,70) {\textcolor{black}{$t=72$}}
        \end{overpic}
    \end{subfigure}
    \begin{subfigure}[c]{.24\textwidth}
        \begin{overpic}[trim={8cm 4cm 8cm 4cm},clip,width=\textwidth]{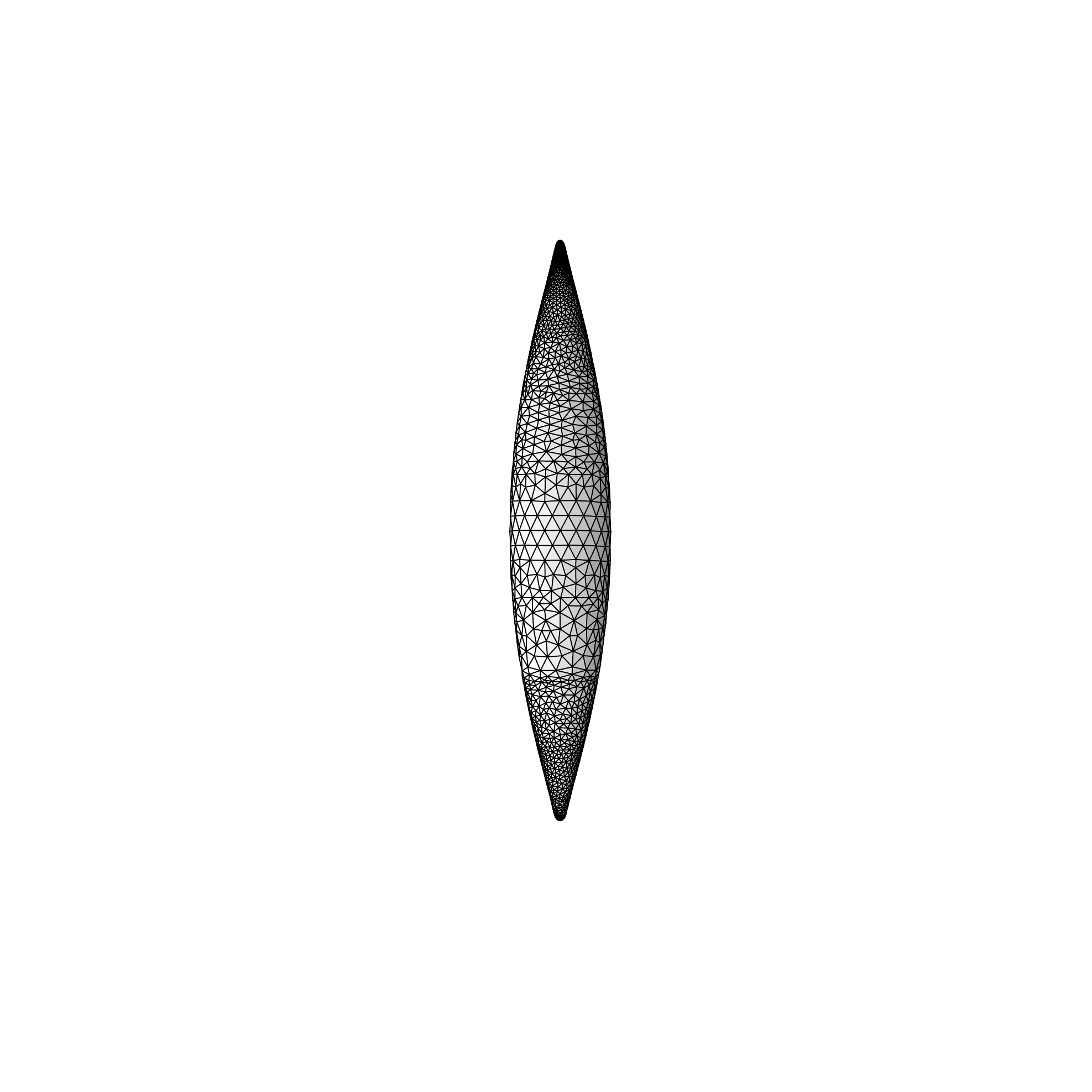}
            \put (0,70) {\textcolor{black}{$t=81$}}
        \end{overpic}
    \end{subfigure}
    \caption{Droplet stretching in a constant field with $Bm=5,\ \mu=30,\ \lambda=10$. Conical tip development can be observed as well as node addition in regions of high surface curvature. Simulation was stopped at $t=80.96$ before reaching the equilibrium value due to prohibitively many points near the tips of the droplets.}
    \label{fig:constant_stretching}
\end{figure}

\begin{figure}[h]
\centering
\begin{minipage}[t]{0.47\textwidth}
    \centering
    \includegraphics[trim={0 0 0 0cm},clip,width=\textwidth]{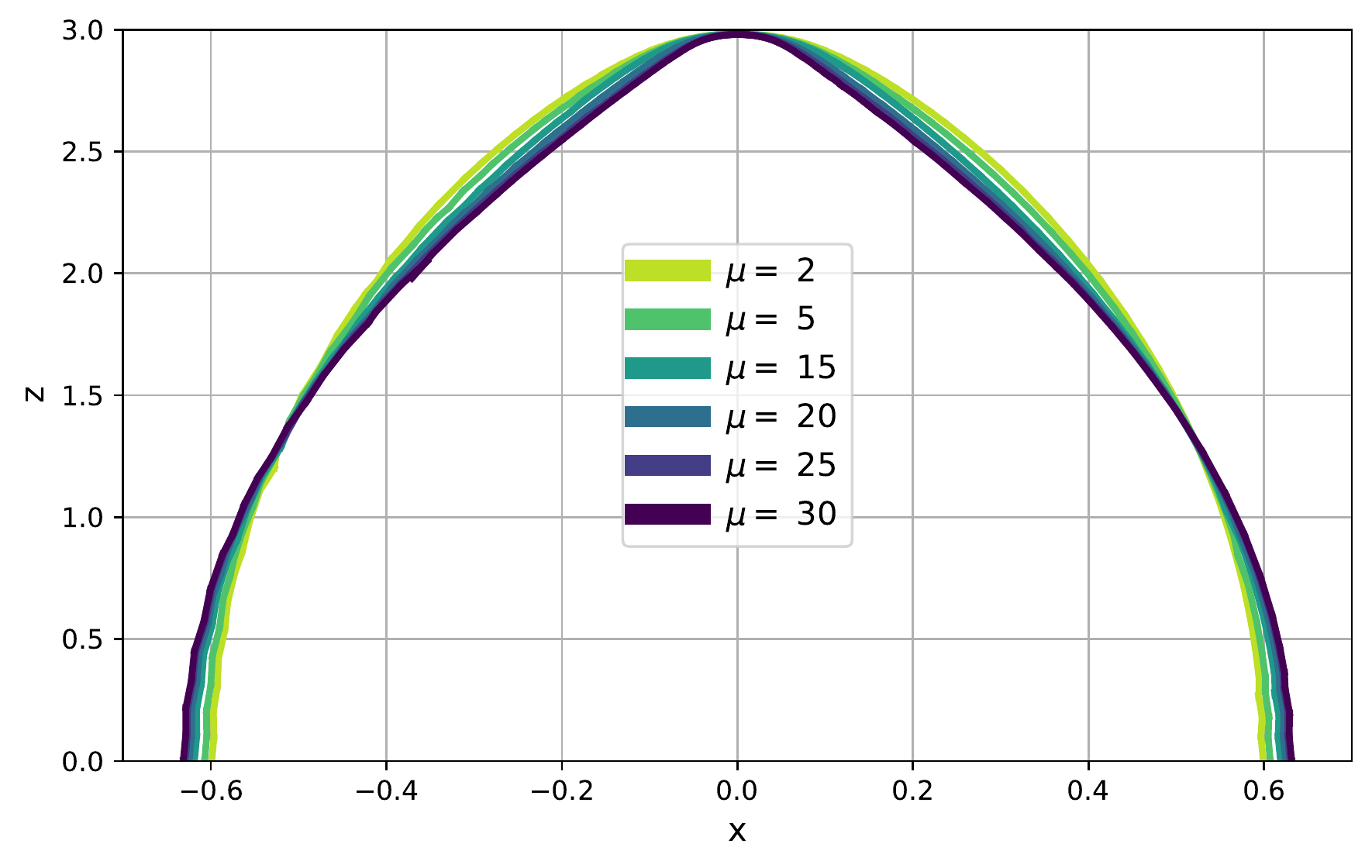}
    \caption{Outlines of droplets that are being elongated at various $\mu$ values in a constant magnetic field, obtained by projecting the full 3D simulation in a plane. All droplets are being elongated in a field that corresponds to an equilibrium axis ratio of $a/b=13$, the graph shows a snapshot when their tips reached $z=3$. It is observed that at larger $\mu$ values the droplets develop sharper tips even when the equilibrium shape has not yet been reached. The axes have been rescaled to emphasize differences between the outlines.}
    \label{fig:hull_comparison}
\end{minipage}
\hspace{1mm}
\begin{minipage}[t]{.49\textwidth}
    \centering
    \includegraphics[width=\textwidth]{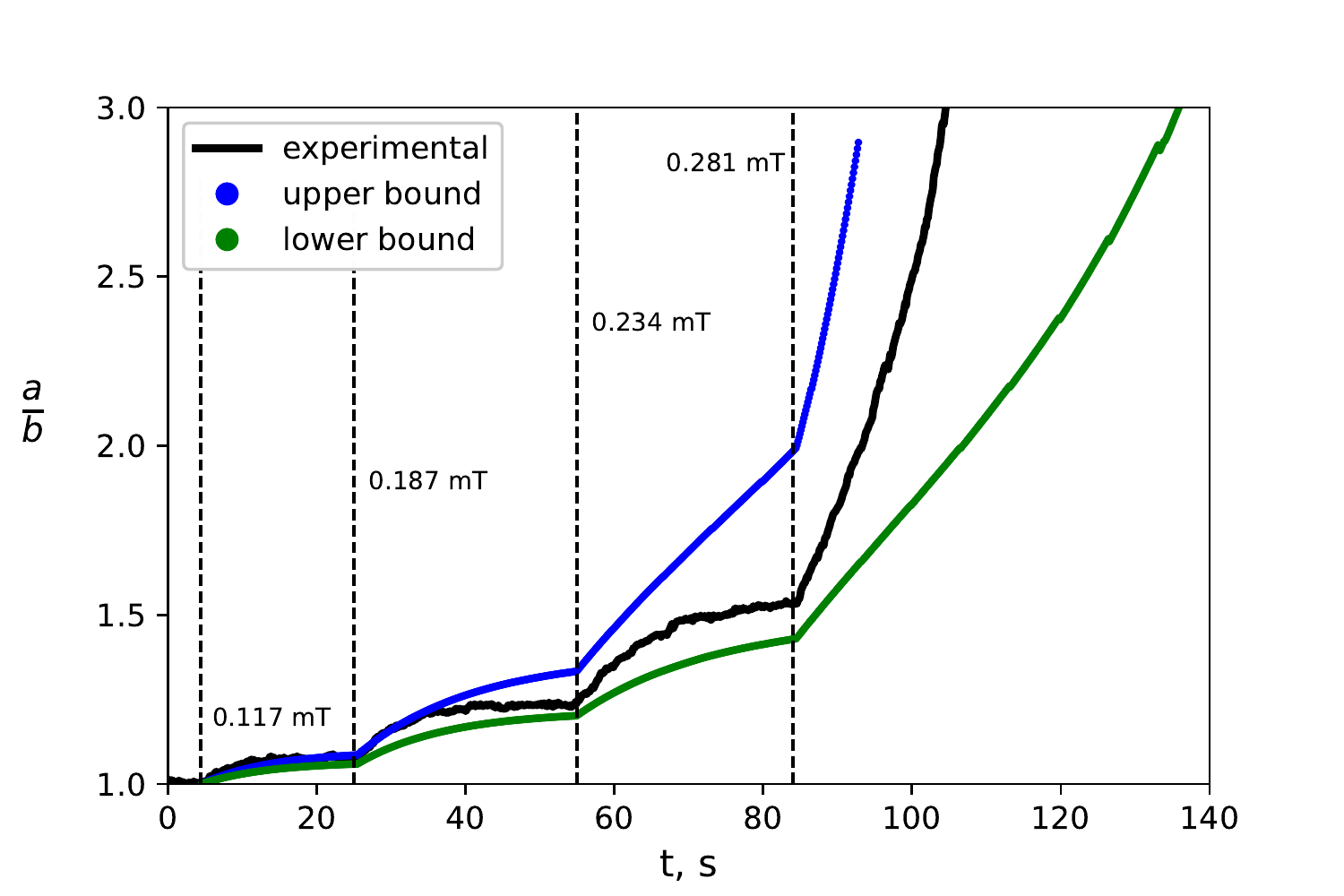}
    \caption{Elongation of a magnetic droplet. Black curve -- experiment with $\mu=34\pm1.5$, $\gamma = (8.2\pm0.4)\cdot10^{-7}J/m^2$, $\lambda=10.1\pm2.5$) in an external field. The blue and green curves are obtained numerically with ($\mu=34$, $\gamma = 7.7\cdot10^{-7}J/m^2$, $\lambda=7.6$) and ($\mu=34$, $\gamma = 8.2\cdot10^{-7}J/m^2$, $\lambda=7.6$), respectively. The magnetic field at first is 0, and is increased at the dotted lines.}
    \label{fig:elongation_exp}
\end{minipage}%
\end{figure}

A spherical magnetic droplet subjected to a constant homogeneous magnetic field elongates at a certain rate until it reaches an equilibrium point where the surface tension balances the magnetic forces and possibly even develops sharp conical tips similar to what has been predicted \cite{stone_drops_1999} and observed \cite{bacri_instability_1982} before, as shown by simulation results in Figure \ref{fig:constant_stretching}.

The Figure \ref{fig:hull_comparison} shows qualitatively distinct behaviour for different values of magnetic permeability $\mu$.
At larger $\mu$ values during the stretching of the droplet, it may develop conical tips, an effect not observed at lower $\mu$ values \cite{stone_drops_1999}. 

Such conical tip development has been captured previously in axi-symmetric simulations \cite{lavrova_numerical_2006}.

It is possible to capture the rate of droplet elongation using this algorithm. Figure \ref{fig:elongation_exp} shows experimentally the elongation of a magnetic droplet with step-wise increasing magnetic field. 
The characteristics of the droplet were determined by an elongation--relaxation measurement of the droplet as described in \citep{erdmanis_magnetic_2017}, and are stated in Figure \ref{fig:elongation_exp}. 
By varying the droplet parameters within the margins of error, we were able to calculate rates of elongation significantly faster and slower than the ones observed experimentally. 
Therefore, by finding the elongation rate that best fits the observation it should be possible to notably increase the precision in the measurements of droplet parameters. 
This is something not easily done for microscopic droplets. 

It is, however, worth mentioning that the calculations are significantly time consuming and there are three parameters that should be honed in on (the viscosity ratio $\lambda$, surface tension coefficient $\gamma$ and the relative magnetic permeability $\mu$), which makes using this method for determining them a bit unwieldy at present.
It could perhaps be mitigated by observing only small deformations of the droplet where the simulations could be run with a relatively small number of collocation points.

\subsection{Rotating field}

\subsubsection{Phase plot of different field strengths and frequencies}
\begin{figure}[H]
    \begin{tikzpicture}
        \node[anchor=south west,inner sep=0] at (0,0) {
        \begin{overpic}[width=\textwidth]{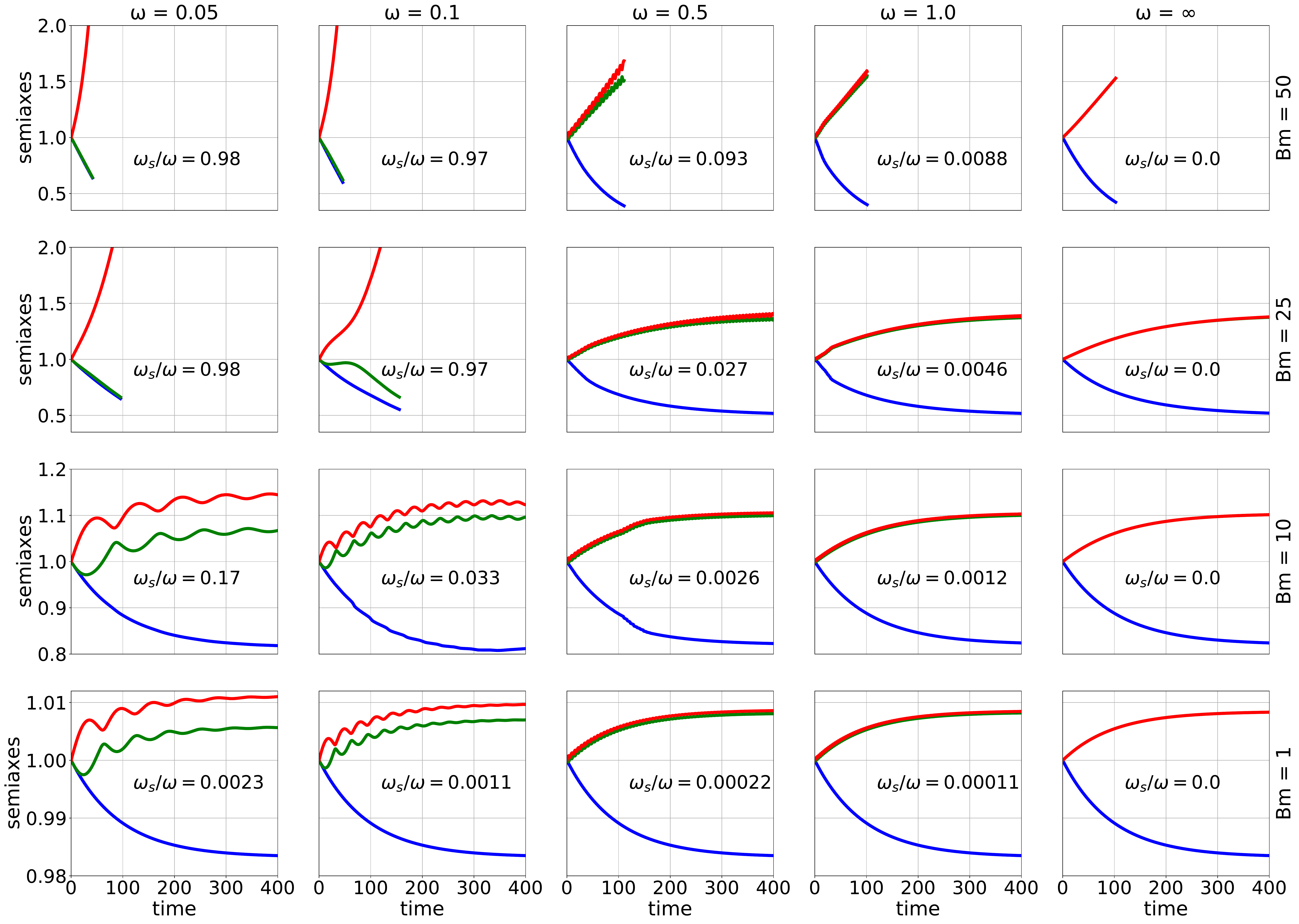}
        \end{overpic}
        };
        \draw[red,ultra thick,rounded corners] (0.1,8.7) rectangle (5.1,4.4);
    \end{tikzpicture}
    \caption{Phase diagram of droplet semi-axes evolution in dimensionless time $t$ in a rotating magnetic field of different dimensionless frequencies $\omega$ and magnetic Bond numbers $Bm$. The region encircled by the red line is shows the region of the phase space where the droplets are elongated, in the rest of the phase space they are flattened. $\omega_s/\omega$ is the ratio of the average surface angular velocity over the magnetic field angular velocity on the last simulation step ($\omega_s = \left< | \vect{r_p} \times \vect{v} | / r_p^2\right>$, $ \vect{r_p}$ is the radius vector projection in the rotation plane). $\omega_s/\omega\approx1$ indicates that the droplet rotates similar to a rigid body, whereas $\omega_s/\omega\ll 1$ indicates, that the droplet seems to rotate following the field, but this apparent rotation is mostly caused by a surface deformation. The droplet parameters are $\lambda=100$, $\mu=10$.}
    \label{fig:rotating_phase_space}
\end{figure}

It has been experimentally observed that magnetic droplets form various qualitatively different shapes in a rotating magnetic field depending on the field frequency and the field strength (Figure 4 in \cite{janiaud_spinning_2000}). 
To explore it numerically, we simulated initially spherical droplets that were subjected to various magnetic field strengths and rotation frequencies (characterized by $\omega$ and $Bm$, respectively), that we show in a phase diagram (Figure \ref{fig:rotating_phase_space}). 
The simulation parameters were $\lambda=100$ and $\mu=10$.

\begin{figure}
    \centering
        \begin{subfigure}[c]{.32\textwidth}
        \begin{overpic}[width=\textwidth,trim={4cm 3cm 5cm 3cm},clip]{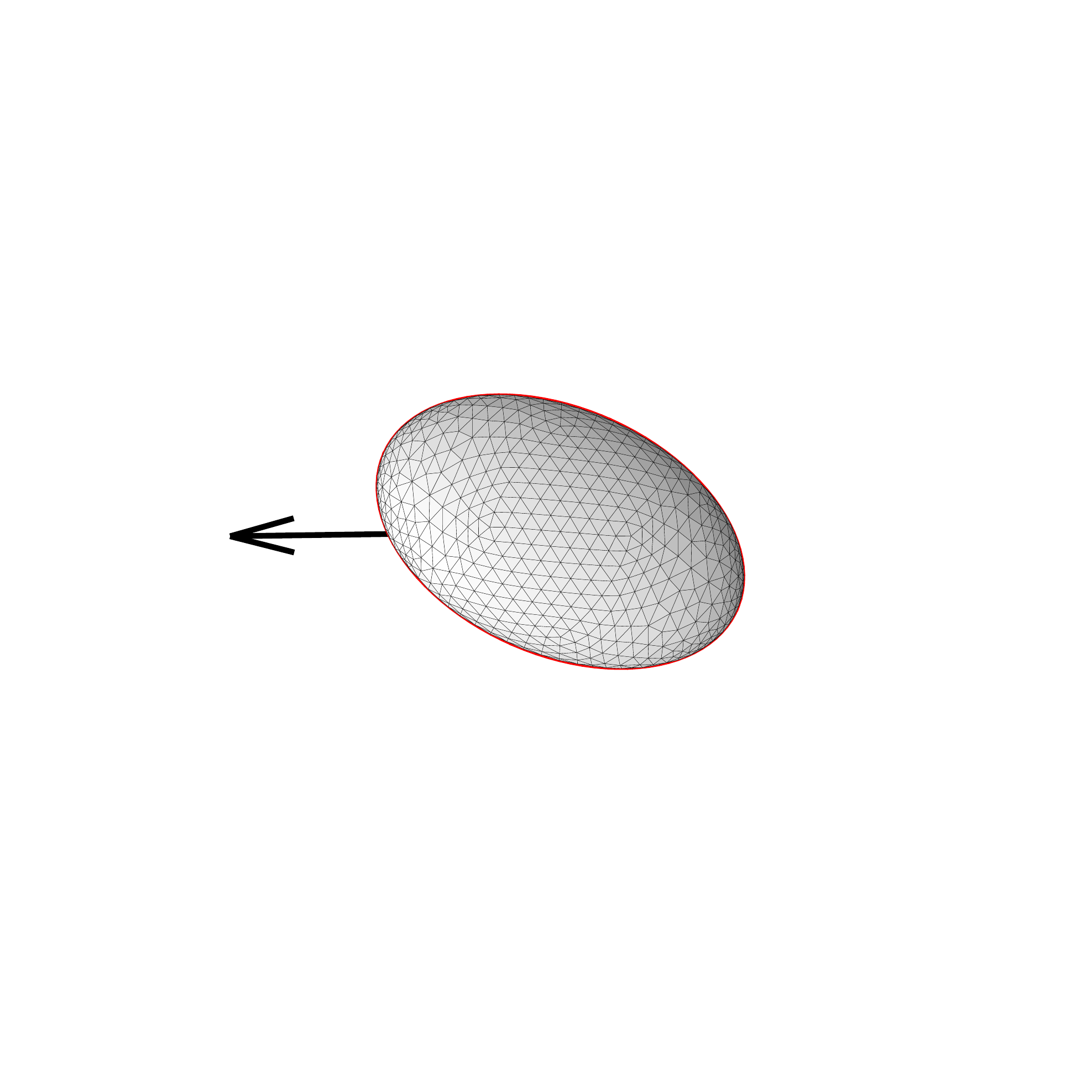}
            \put (0,100) {\textcolor{black}{a)}}
            \put (25,100) {\textcolor{black}{ $t=79$ }}
        \end{overpic}
    \end{subfigure}
    \begin{subfigure}[c]{.32\textwidth}
        \begin{overpic}[width=\textwidth,trim={4cm 3cm 5cm 3cm},clip]{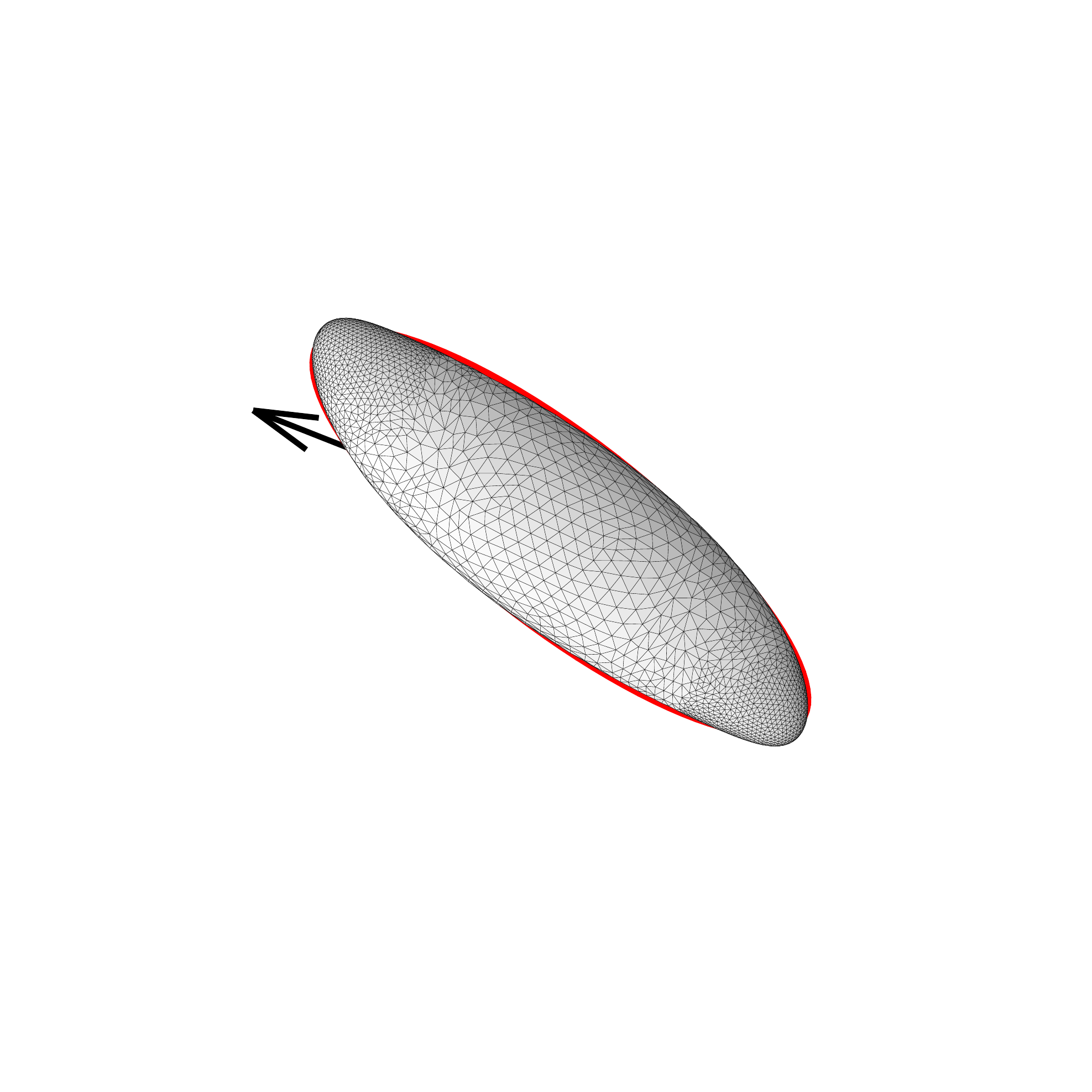}
            \put (25,100) {\textcolor{black}{ $t=138$ }}
        \end{overpic}
    \end{subfigure}
    \begin{subfigure}[c]{.32\textwidth}
        \begin{overpic}[width=\textwidth,trim={4cm 3cm 5cm 3cm},clip]{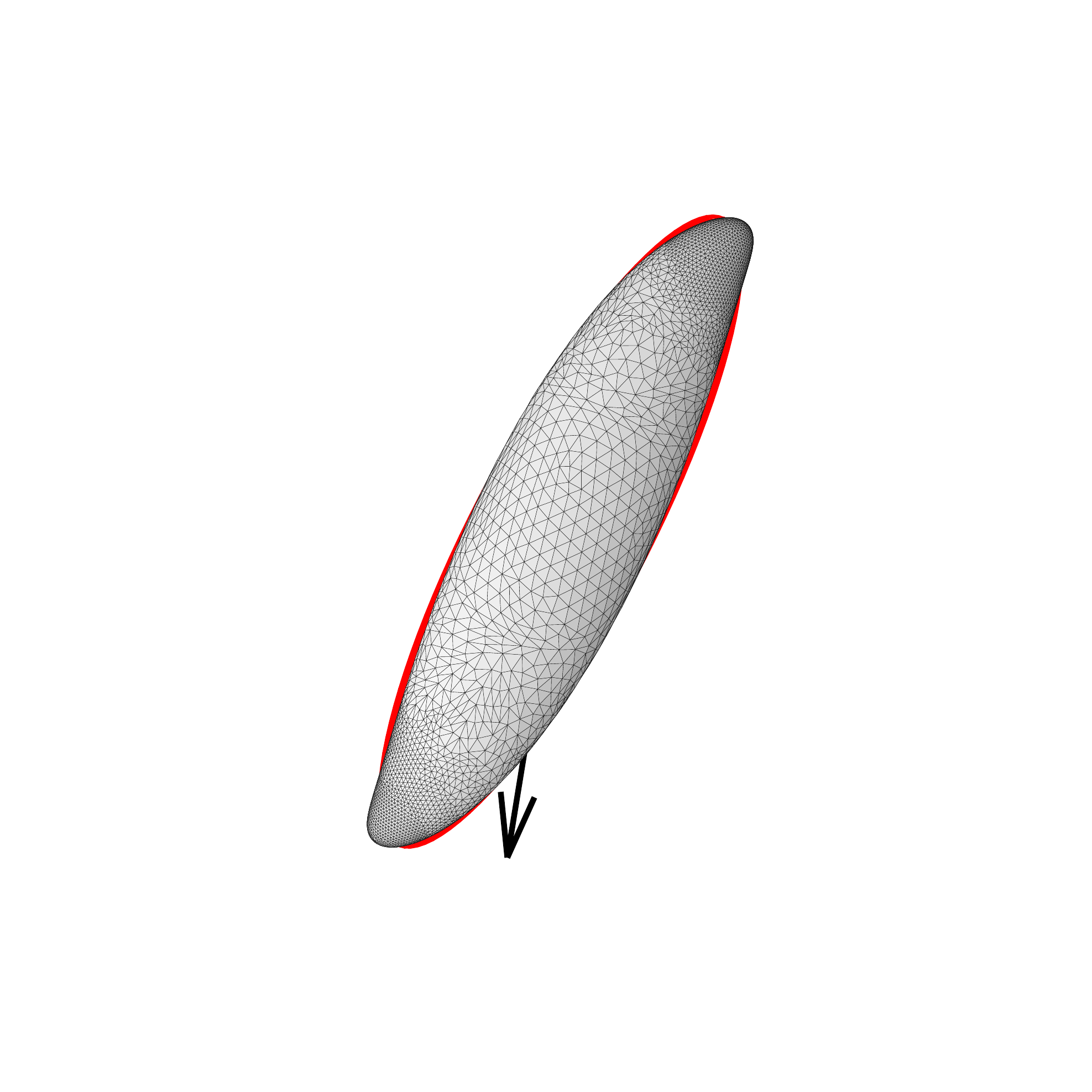}
            \put (25,100) {\textcolor{black}{ $t=155$ }}
        \end{overpic}
    \end{subfigure}
    \begin{subfigure}[c]{0.8\textwidth}
        \begin{overpic}[width=\textwidth,trim={3.5cm 8cm 10cm 8cm},clip]{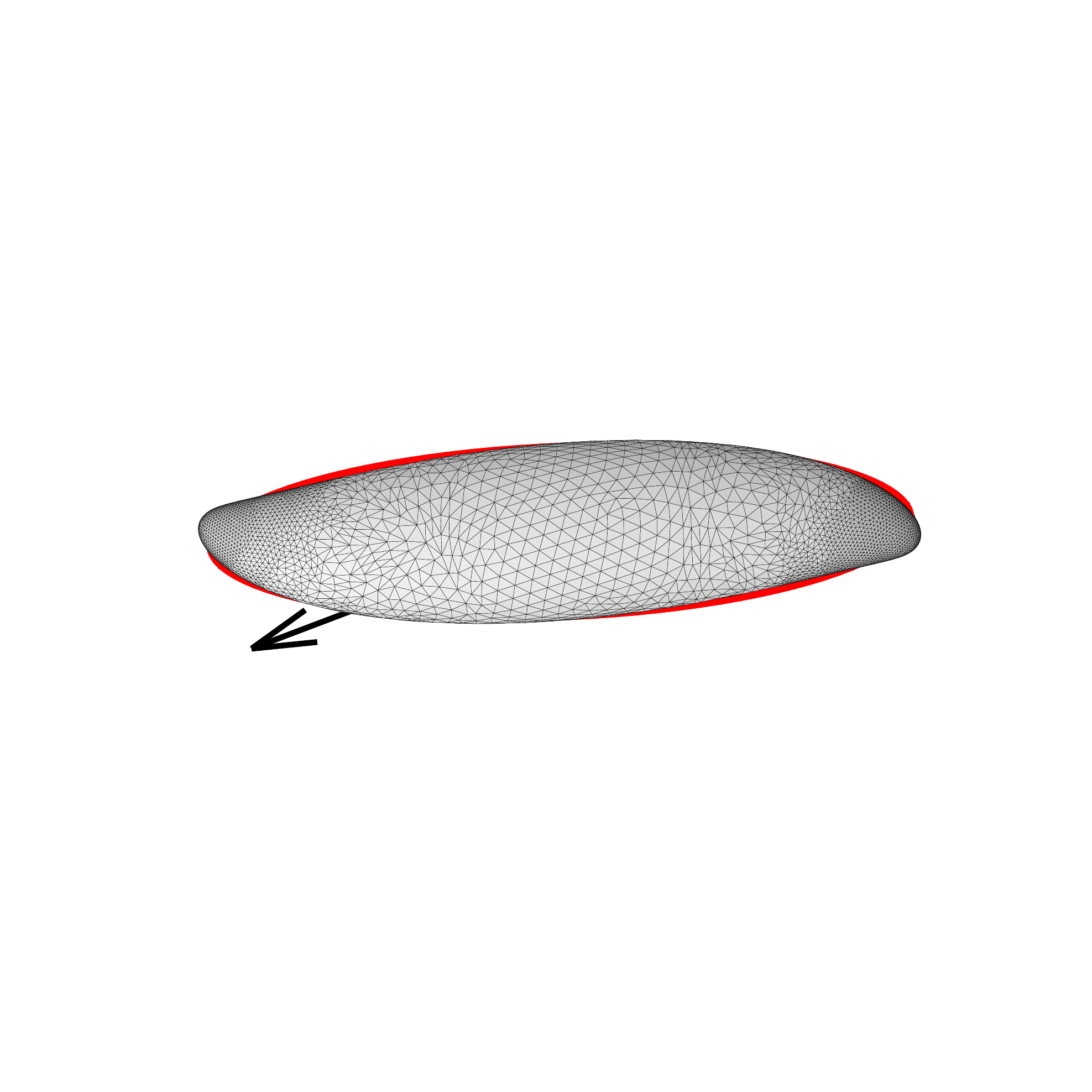}
            \put (-10,60) {\textcolor{black}{b)}}
        \end{overpic}
    \end{subfigure}
    \caption{ a) Shows the evolution of a magnetic droplet in a counter-clockwise rotating magnetic field. The simulation parameters are such that the droplet elongates instead of remaining flattened: $Bm=25$, $\mu=30$, $\lambda=100$, $\omega=0.1$. The mesh is overlaid over a best-fit ellipsoid colored in red. The black arrows indicate the momentary magnetic field direction. b) shows a zoom of part of the droplet at $t=155$ to illustrate the discrepancies between the droplet shape and the best-fit ellipsoid. Notably the tips start deforming in an ``S" shape due to the viscous drag.}
    \label{fig:rotating_mesh}
\end{figure}

We fitted a tri-axial ellipsoid to the points that describe the droplet's surface to show how the semiaxes evolve over time.
It was found that such a fit well described the droplets when the elongation was not too large.
For the droplets inside the red outline in Figure \ref{fig:rotating_phase_space} the fit did not accurately capture the shape of the tips, that started to form more of an ``S" shape as the viscous forces drag them behind (Figure \ref{fig:rotating_mesh}).
For low $\omega$ and $Bm$ the droplets can be described as a tri-axial ellipsoid. 
The shape become more oblate as $\omega$ increases, at $\omega=1$ the shape is already nearly identical to the case when $\omega=\infty$, which is completely axisymmetric.

Similarly to the experimental results \cite{janiaud_spinning_2000}, we also obtained that for low enough $\omega$ and high enough $Bm$ the droplets elongate. 
However, experimentally it was also observed that the field rotation suppresses the elongation and droplets remained spherical.
As can be seen by our simulations, that is not the case.
The droplets take up a quasi-oblate shapes, which observed from the field rotation direction would appear nearly indistinguishable from a spherical droplet. 

\begin{figure}
    \centering
    \begin{subfigure}[c]{.49\textwidth}
        \begin{overpic}[width=\textwidth]{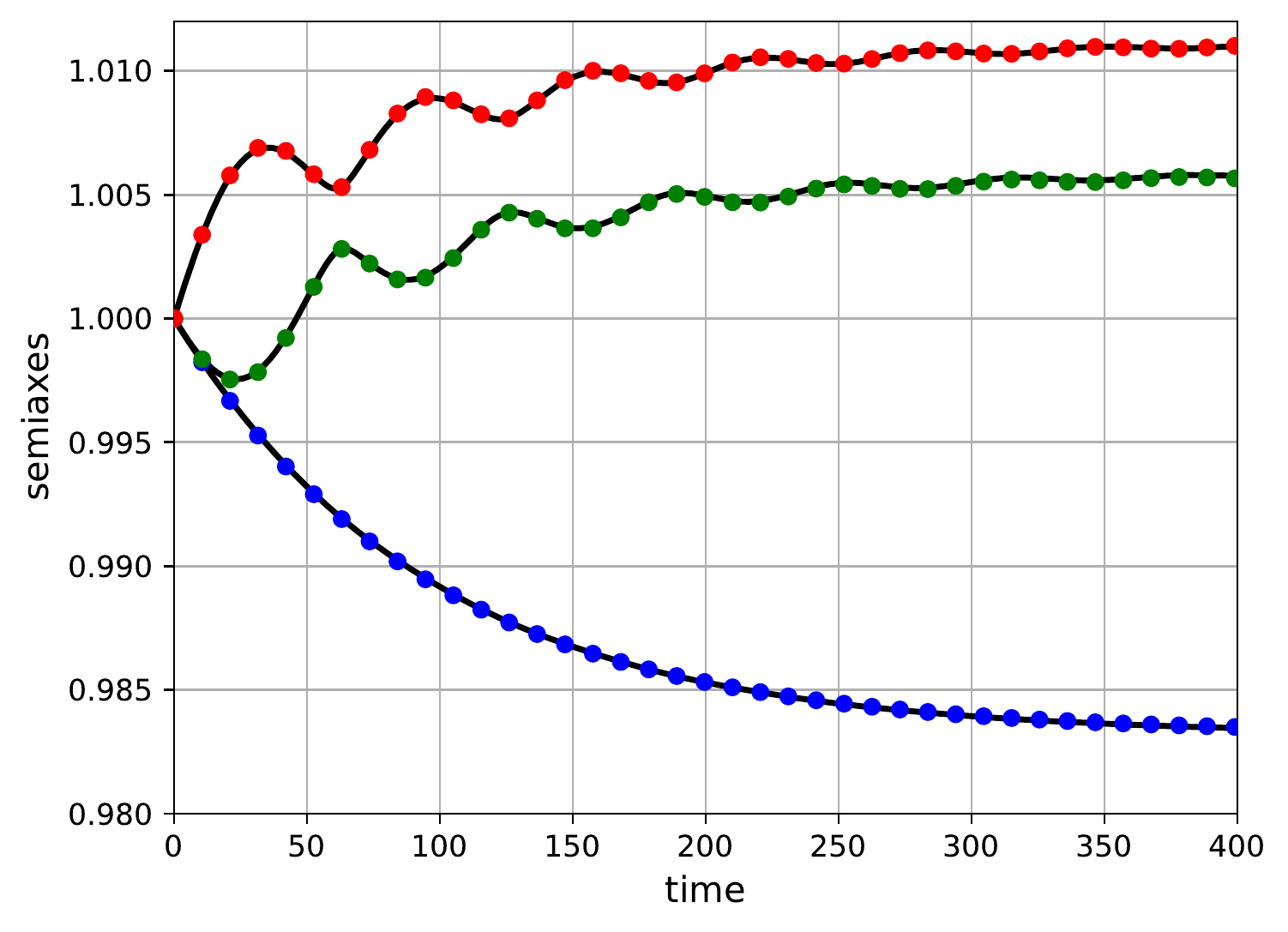}
            \put (0,70) {\textcolor{black}{a)}}
        \end{overpic}
    \end{subfigure}
    \begin{subfigure}[c]{.49\textwidth}
        \begin{overpic}[width=\textwidth]{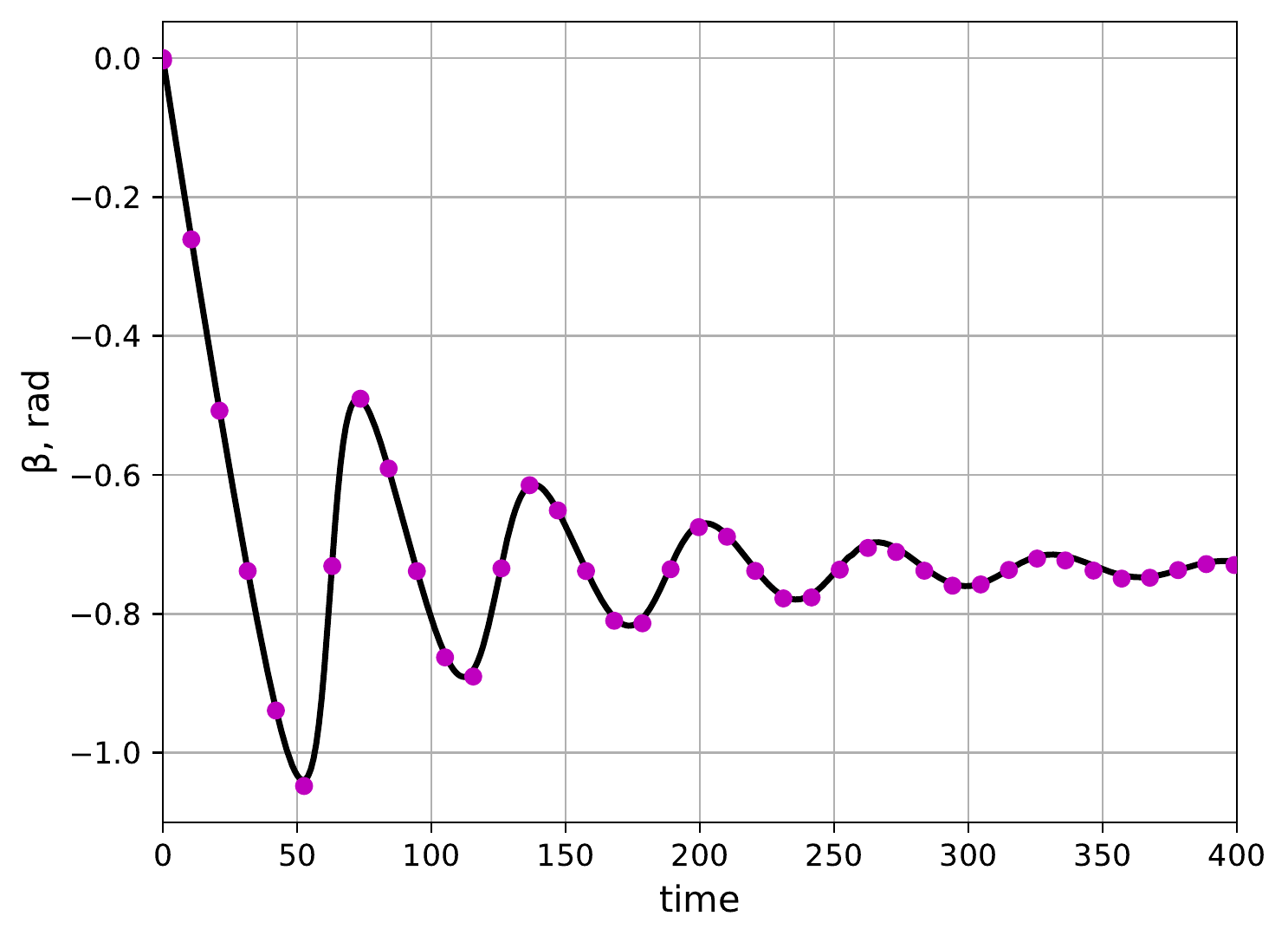}
            \put (0,70) {\textcolor{black}{b)}}
        \end{overpic}
    \end{subfigure}
    \caption{The evolution of an initially spherical droplet's shape in a rotating magnetic field. The points are the simulation results (every 100th is shown for clarity) and the black lines are from the equation \eqref{eq:eps_system}. The simulation parameters are $\lambda=100$, $\mu=10$, $Bm=1$ and $\omega=0.05$. a) shows the evolution of the droplet semiaxes and b) shows the angle $\beta$ between the droplet's largest axis and the magnetic field. Negative $\beta$ means that the droplet is trailing the field. }
    \label{fig:small_deform}
\end{figure}

A small deformation theory based on a phenomenological anisotropy tensor formalism was recently developed to describe the motion of magnetic droplets in a rotating field valid up to $O(Bm)$ \cite{stikuts2021small}.
There it is shown that the droplet can be described by an ellipsoid with semiaxes $a \leq b \leq c$ and an angle $\beta$ between the droplet's largest axis and the magnetic field, such that $\beta<0$ if the droplet is trailing the field.
The droplet's shape evolution is governed by 
\begin{equation}
\left\{ \begin{aligned} 
  \frac{d \epsilon_1}{d t} &= -\frac{1}{\tau} \left( \epsilon_1 - \delta \cos(2\beta) \right) \\
  \frac{d \epsilon_2}{d t} &= -\frac{1}{\tau} \left( \epsilon_2 - \delta \sin^2(\beta) \right) \\
  \frac{d \beta}{d t} &= -\omega - \frac{\delta \cos(\beta)\sin(\beta)}{\tau \epsilon_1}
\end{aligned} \right. ,
\label{eq:eps_system}
\end{equation}
where $\epsilon_1=(a-b)/b$, $\epsilon_2=(b-c)/b$ together with the incompressibility condition $abc=1$ fully determine the semiaxes, $\tau$ is the dimensionless small deformation relaxation time \eqref{eq:tau_relax}, $\omega$ is the dimensionless magnetic field rotation angular frequency and
\begin{equation}
    \delta = \frac{9 Bm}{32\pi} \frac{\left(\mu-1\right)^2}{\left(\mu+2\right)^2}.
\end{equation}
For small $Bm$ the BEM simulations follow closely to the shape predicted by \eqref{eq:eps_system} (Figure \ref{fig:small_deform}). 
The simulations were also used to determine the limits of the small deformation theory \citep{stikuts2021small}.

How do the droplets change their orientation in the rotating field?
They can rotate or the change of orientation can arise due to surface deformations.
To determine which of these effects is dominant, we compare the angular velocity of the magnetic field $\omega$ with the average surface angular velocity
\begin{equation}
    \omega_s = \left< \frac{| \vect{r_p} \times \vect{v} |}{r_p^2}\right>,
\label{eq:omega_s}
\end{equation}
where the averaging is done over all the nodes, $\vect{r_p}$ is the radius vector to a given node projected to the rotation plane.

In the region where $\omega$ is small and $Bm$ large (denoted by the red outline in Figure \ref{fig:rotating_phase_space}) the droplets elongate and rotate in the direction of the field more or less like a rigid body ($\omega_s/\omega \approx 1$), albeit still deforming. 
When the $Bm$ is small or $\omega$ large, the droplets take up a flattened shape, which still has one axis larger than the other. 
The flattened drops also seem to rotate in the direction of the field, but this apparent rotation is caused by surface deformations, which can be seen by the fact that the surface of the droplet almost does not rotate ($\omega_s/\omega \ll 1$). 

In the phase space outside the region where droplets become elongated, their axes undergo oscillations with roughly twice the magnetic field frequency as they settle into equilibrium values, these oscillations get smaller in magnitude as $\omega$ increases.
These results agree with the $O(Bm)$ small deformation theory \cite{stikuts2021small}, where it was shown that close to the equilibrium shape the lengths of the droplet axes oscillate with $2\omega$ and the droplet surface angular velocity scales as $O(Bm^2)$.



\subsubsection{Back--and--forth motion}

\begin{figure}[H]
    \centering
    \begin{overpic}[scale=0.6]{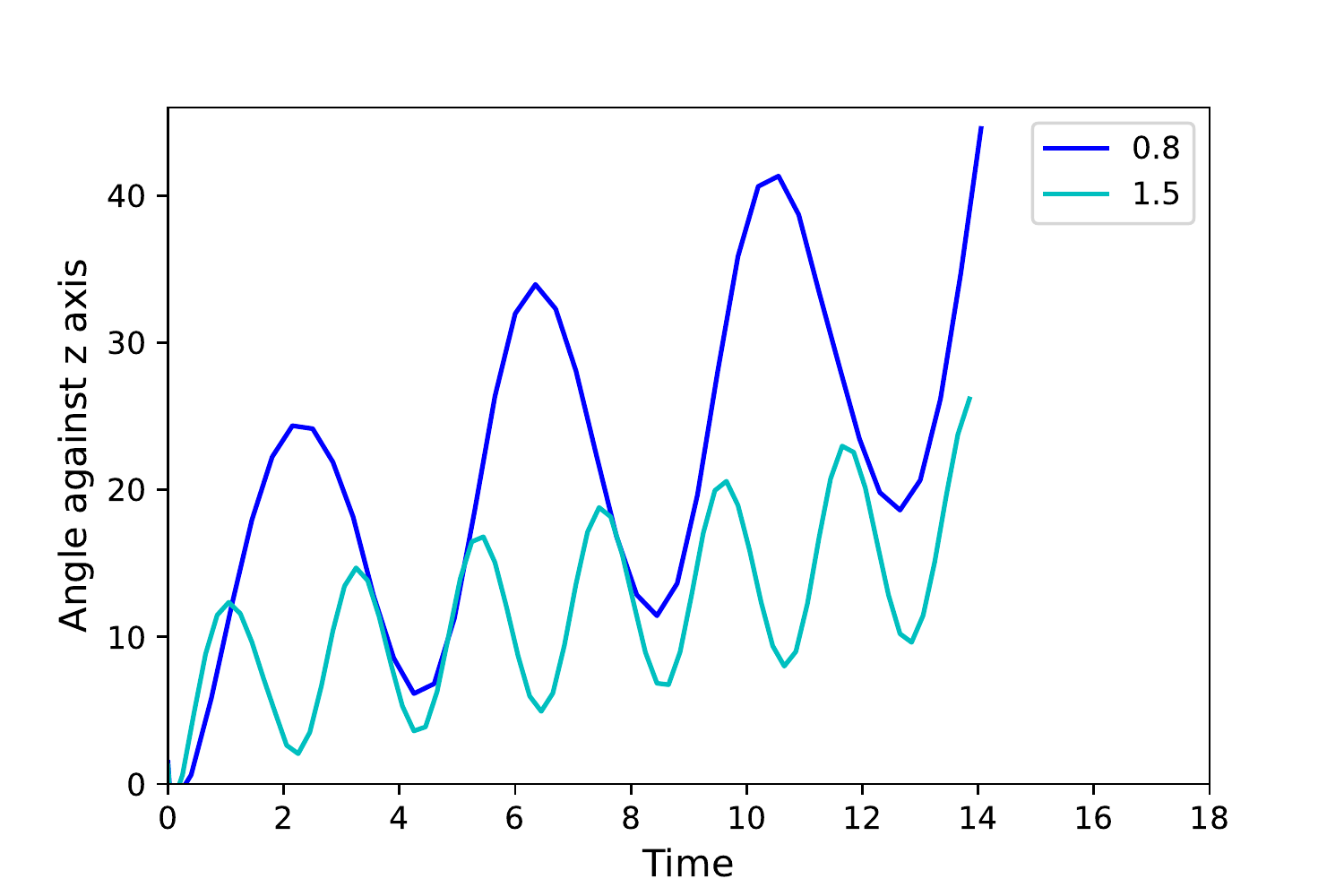}
    \put(74.4,8.4){\frame{\includegraphics[trim={9cm 6cm 9cm 6cm},clip,scale=0.2]{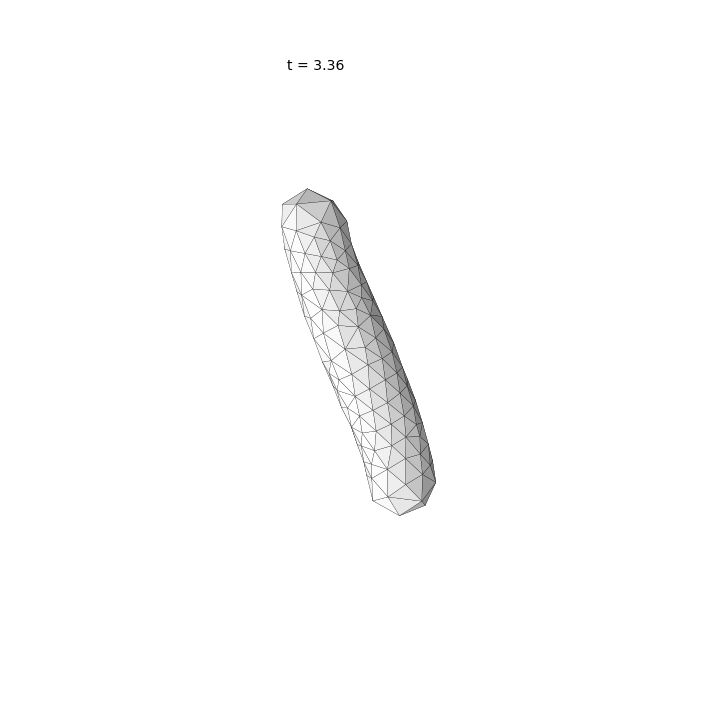}}}
    \end{overpic}
    \caption{Example of observed back--and--forth motion of an initially elongated droplet (shown in the inset) in moderately fast rotating fields of different frequencies $\omega$. The lines indicate the angle between longest axis of the droplet and the stationary $z$ axis. Simulations used $\mu=30,\ \lambda=1$.}
    \label{fig:back_forth}
\end{figure}


Another approach investigated in a rotating field was a droplet, initially extended in a constant field and stretched to an axis ratio of about 7, was put in a rotating field at various frequencies. At low frequencies the droplet just follows the external field.
At higher frequencies a back-and-forth motion was observed as shown in Fig. \ref{fig:back_forth}, similarly, as is the case with a solid elongated paramagnetic particle \cite{cimurs_stability_2019}, magnetotactic bacteria \cite{erglis_dynamics_2007} and self--propelling magnetic particles with a permanent magnetic moment \cite{cebers_dynamics_2006}.
In these simulations node addition was disabled since accurate description of such highly elongated droplets would require prohibitively many points, other mesh maintenance techniques were still employed.
Therefore the results should be interpreted only qualitatively.

\subsection{Field threshold for starfish stability}

\begin{figure}[h]
    \centering
    \includegraphics[trim={0 0 0 2cm},clip,width=0.8\textwidth]{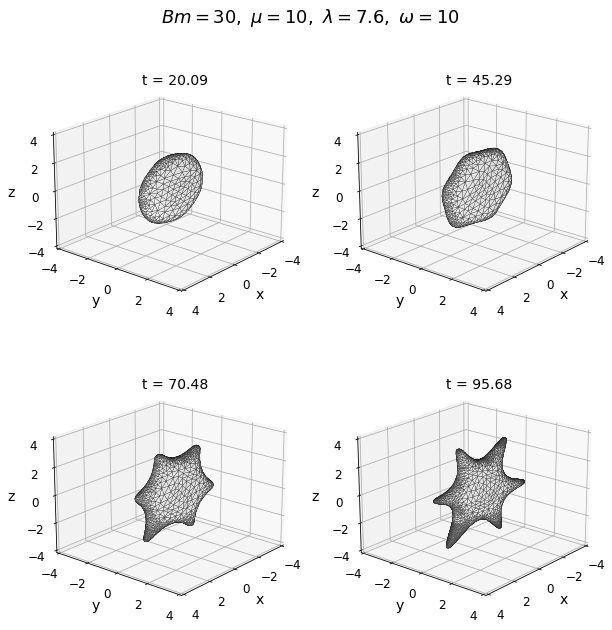}
    \caption{Example of the starfish instability using $Bm=35,\ \mu=10,\ \omega=10,\ \lambda=7.6$ and the node addition cut-off criteria of 0.4.}
    \label{fig:starfish_evolution}
\end{figure}

\begin{figure}[h]
    \centering
    \includegraphics[width=0.9\textwidth]{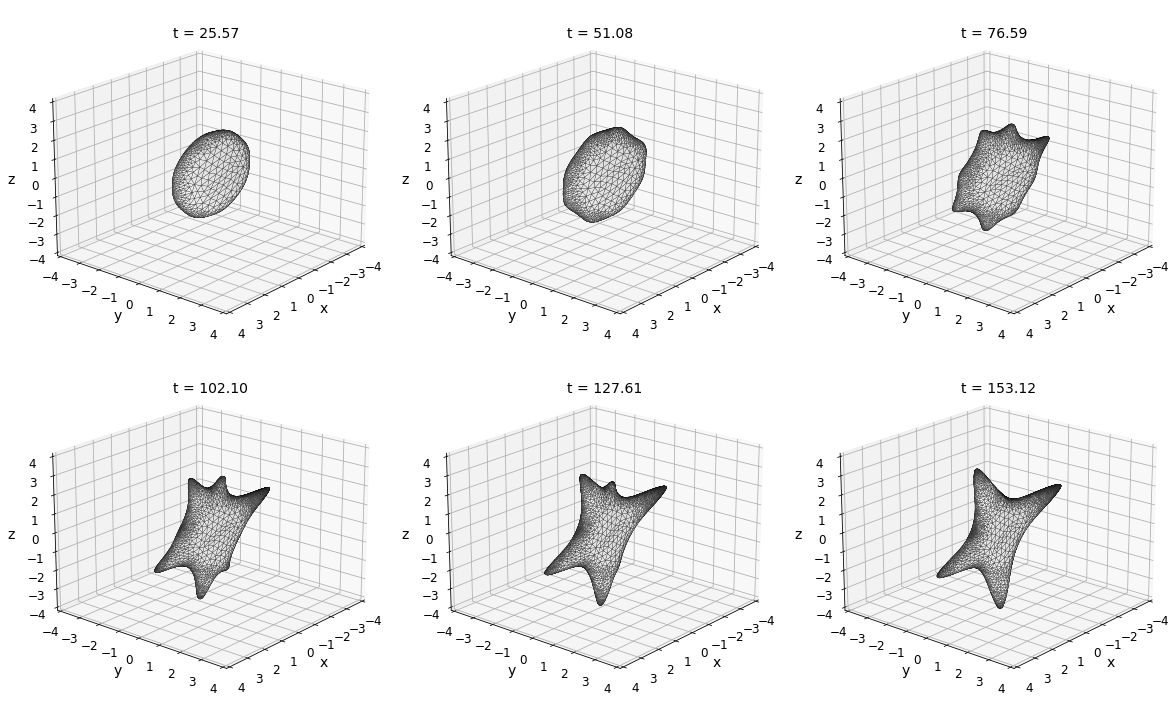}
    \caption{Example of various starfish mode competition using $Bm=30,\ \mu=10,\ \omega=~10,\ \lambda=7.6$ and the node addition cut-off criteria of 0.4. At first six tips seem to be forming, but at a later time only four remain.}
    \label{fig:starfish_modes}
\end{figure}

The algorithm allows to investigate the onset of the starfish instability shown in Figure \ref{fig:starfish_evolution}, known to occur at strong enough rotating and fast enough magnetic fields, whereby the oblate droplet develops finger--like structures on its perimeter \cite{bacri_behavior_1994}.
We have observed competition between these modes (configurations of different number of ``fingers''), as shown in Figure \ref{fig:starfish_modes} where the initially manifested mode of $n=6$ is subsequently overtaken by the mode $n=4$.
In both of these cases the droplet developed these ``fingers'' spontaneously from some asymmetries in the droplet mesh.

To investigate carefully the evolution and competition of different modes, we start with a axially symmetric oblate ellipsoid, obtained from a minimal energy configuration in an infinitely fast rotating field \cite{bacri_behavior_1994, morozov_bifurcations_2000}, with an added small initial sinusoidal perturbation of a specific mode $n$ (representing the number of starfish ``fingers'') along its perimeter in the plane of the rotating field $r(\theta) = \varepsilon \cos{(\omega_n \theta)}$ with $\varepsilon = 0.01 R_0$. 
The perturbed droplet is then allowed to evolve in an period-averaged high frequency field (the governing equations for the averaged field are described in \cite{erdmanis_magnetic_2017}) of varying strengths.

\begin{figure}[H]
\centering
\begin{minipage}[t]{.47\textwidth}
    \centering
    \includegraphics[width=0.97\textwidth]{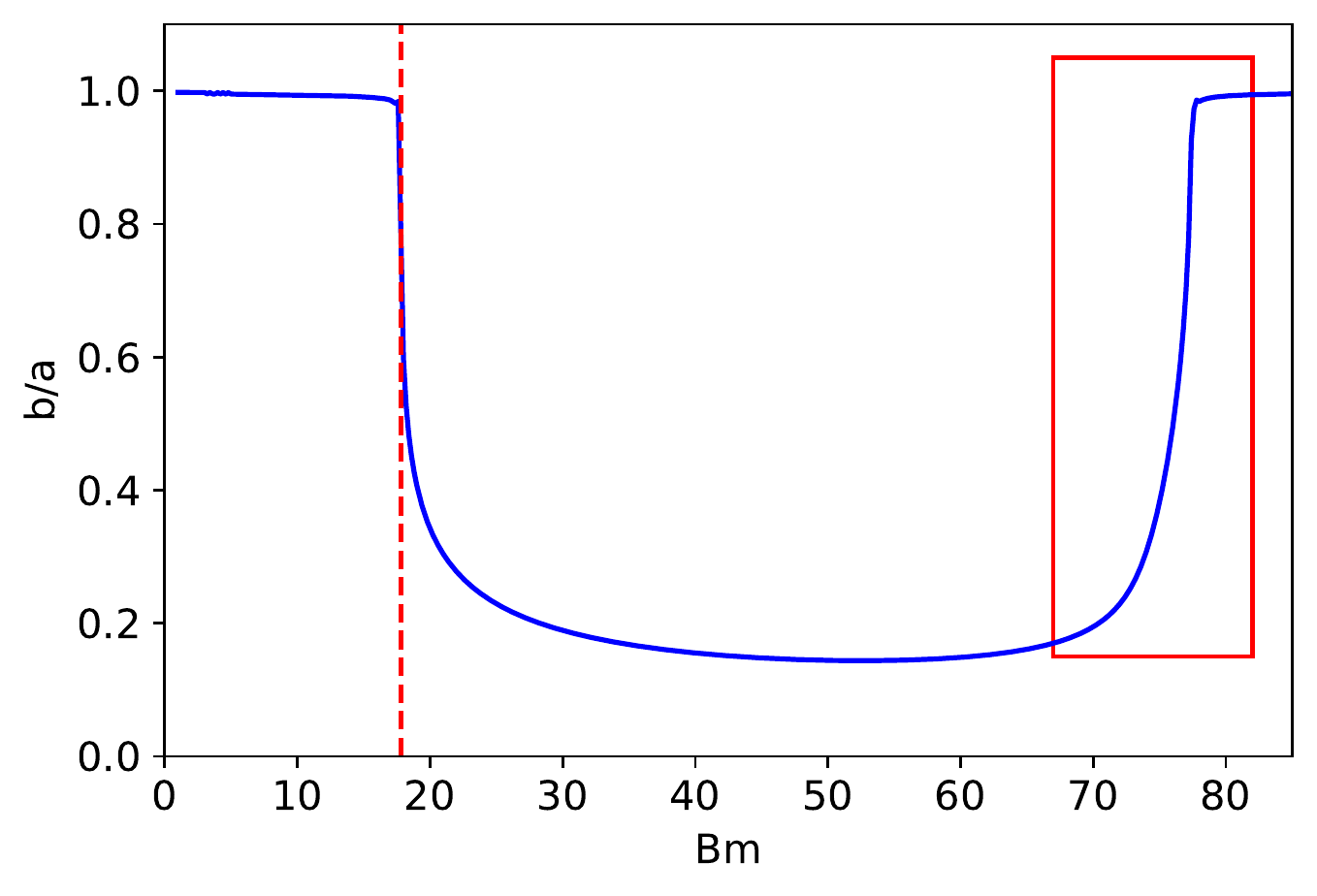}
    \caption{The vertical line on the left indicates the critical field determined by analysis of the logarithmic increment $\beta_2$, it coincides very well with the value determined via droplet energy minimization. The region on the right represents the droplet becoming oblate again.}
    \label{fig:ba_vs_Bm}
\end{minipage}%
\hspace{1mm}
\begin{minipage}[t]{.47\textwidth}
    \centering
    \includegraphics[width=\textwidth]{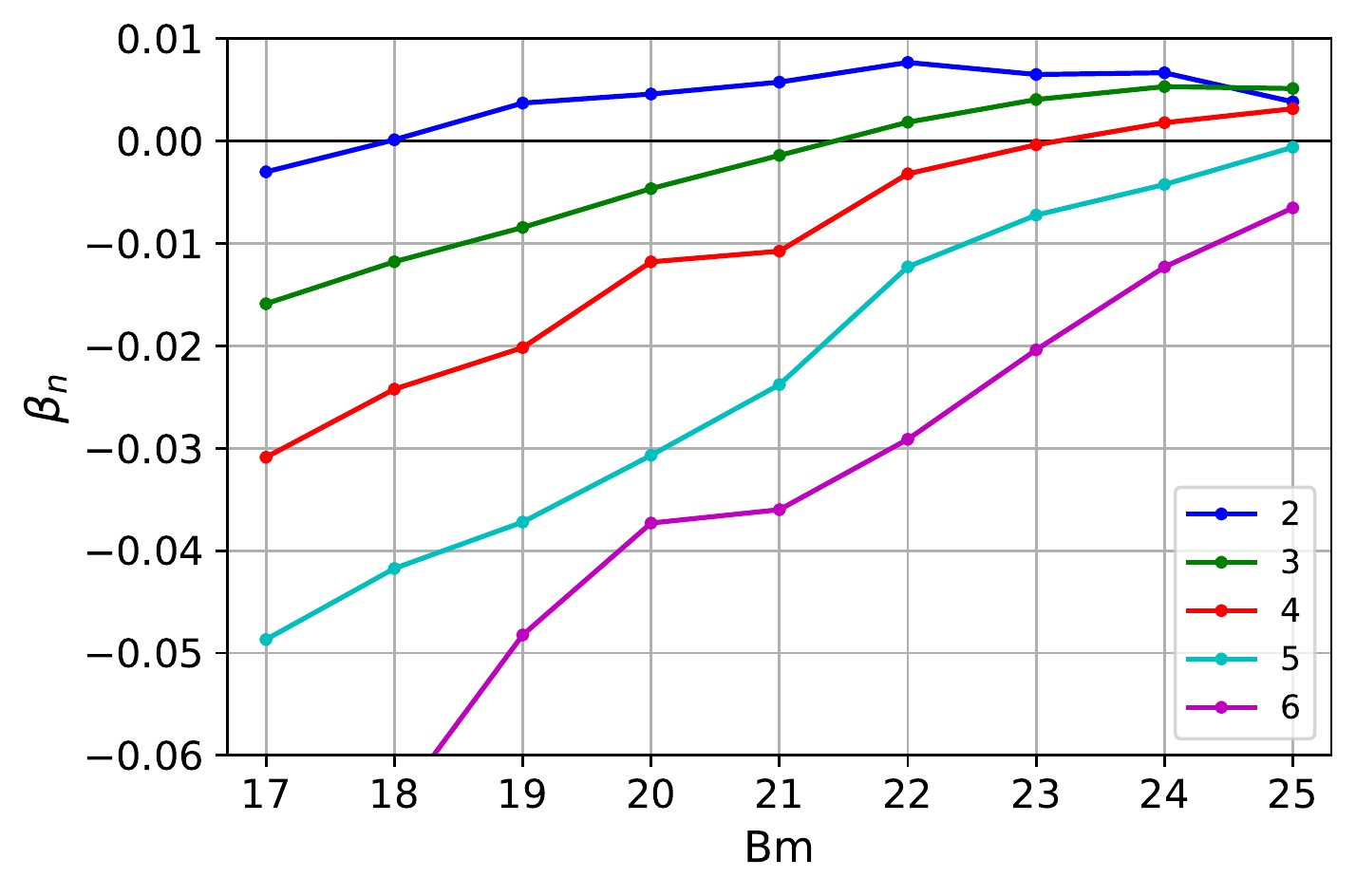}
    \caption{Rates of growth $\beta_n$ of various ``finger'' modes at different magnetic fields $Bm$. Competition of modes can be observed as the logarithmic increments $\beta$ approach one another at larger magnetic fields.}
    \label{fig:starfish_beta}
\end{minipage}
\end{figure}

One goal aim of such an analysis would be to determine the critical magnetic field value at which the oblate-prolate transition occurs, shown in Figure \ref{fig:ba_vs_Bm}, as indicated by the onset of the $n=2$ mode.
Finally, analysis of Fourier amplitudes of the droplet's perimeter yields the dynamics of the mode amplitude as shown in Figure \ref{fig:starfish_beta}, assuming exponential growth of $n$--th mode near equilibrium ${A_n(t) = C_n e^{\beta_n t}}$ with $\beta_n$ being the logarithmic increment to be determined.

\begin{figure}[H]
\centering
\begin{minipage}[t]{.49\textwidth}
    \centering
    \includegraphics[width=0.94\textwidth]{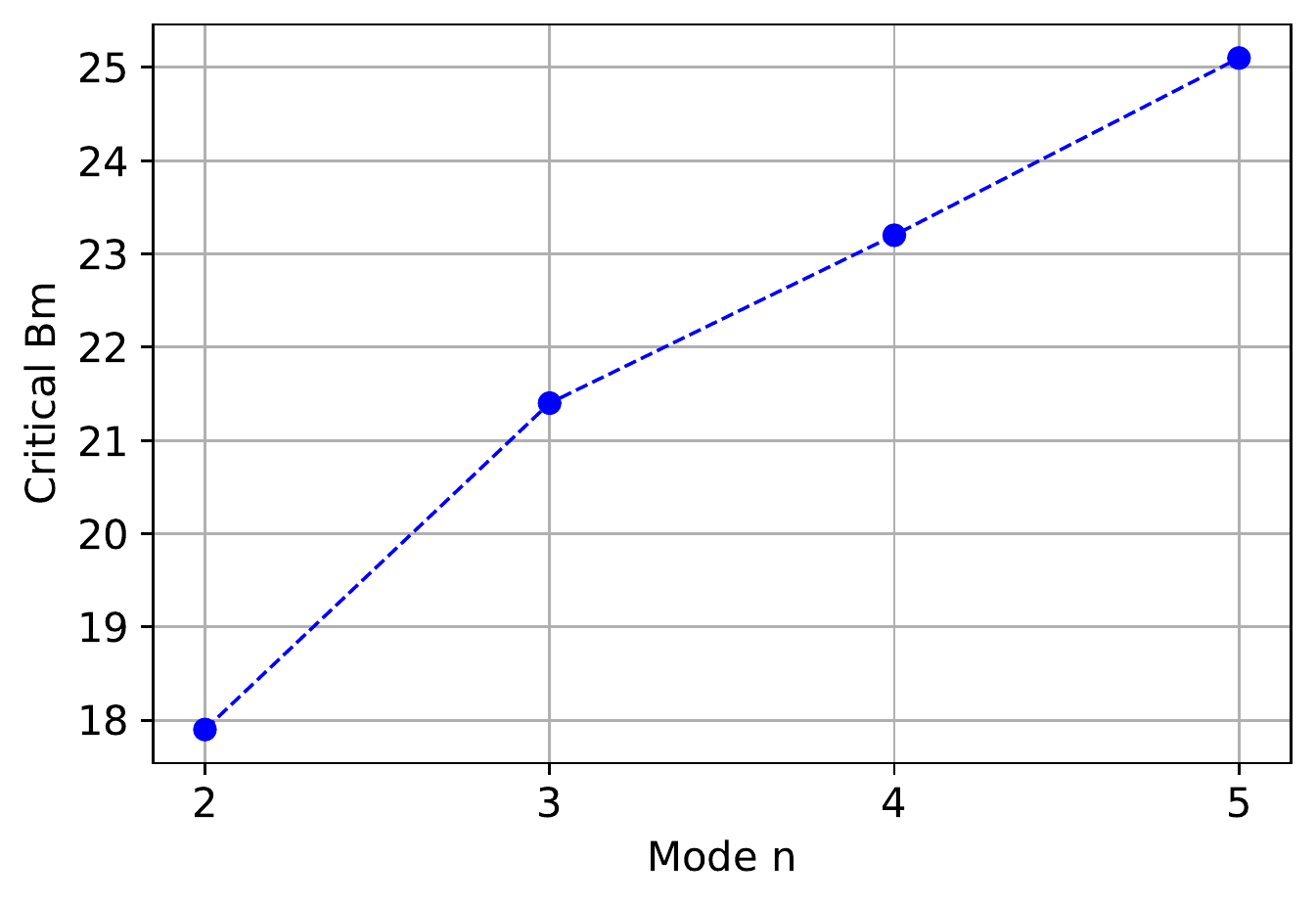}
    \caption{The critical field above which a mode can start growing increases with mode number $n$. This does not mean however that the particular mode will be observed as others might be growing faster.}
    \label{fig:critical_Bm_vs_n}
\end{minipage}%
\hspace{1mm}
\begin{minipage}[t]{.49\textwidth}
    \centering
    \includegraphics[width=\textwidth]{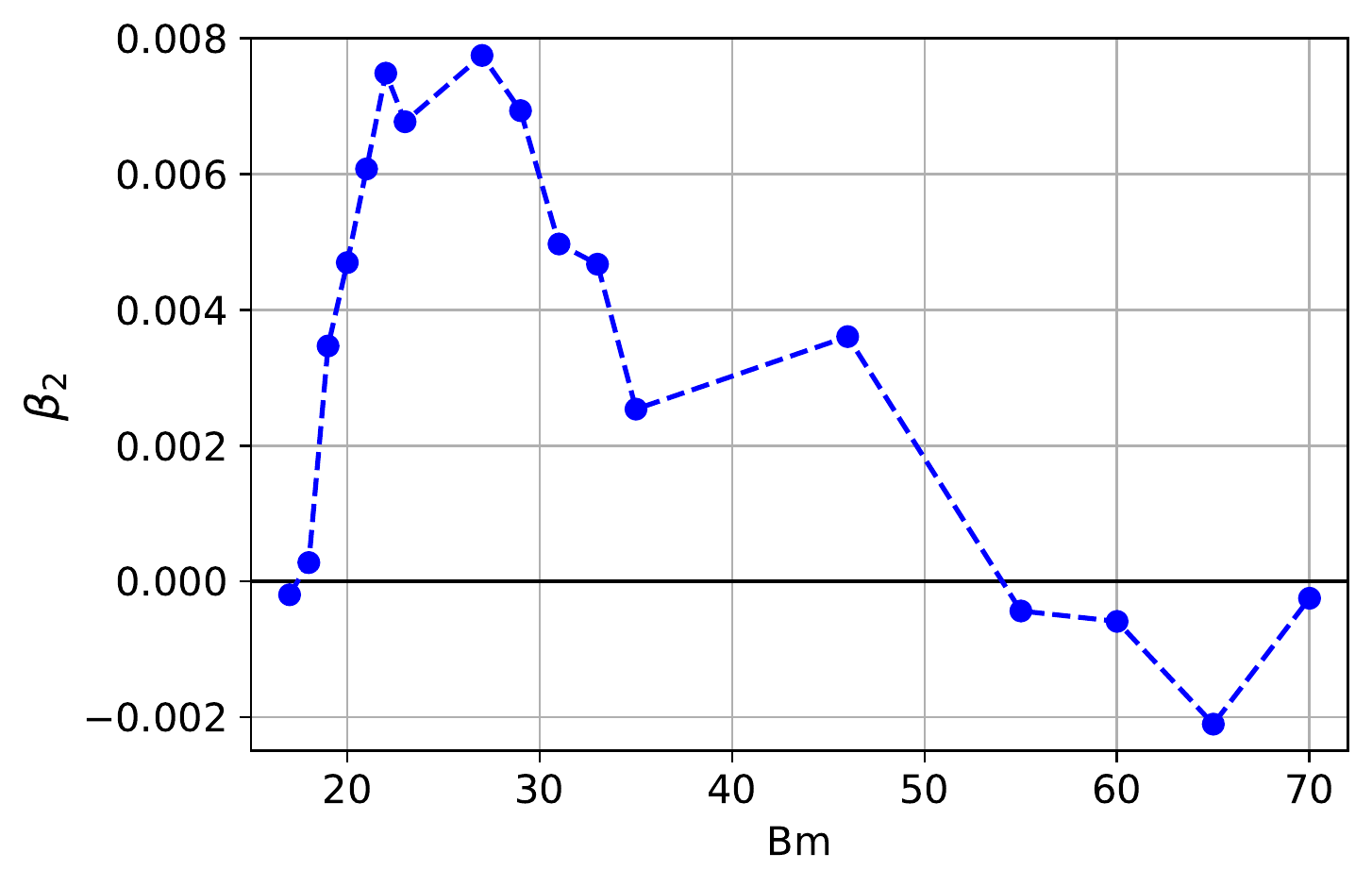}
    \caption{The logarithmic increment of the starfish mode $n=2$ becomes positive in a certain field strength region, indicating a transition to a prolate shape, which can be reversed at large fields, when the droplet becomes oblate again -- this reentrant transition was predicted and observed in \cite{bacri_behavior_1994}.}
    \label{fig:beta2_vs_Bm}
\end{minipage}
\end{figure}

It is observed that the $\beta_n$'s tend to increase with increasing magnetic field strengths, shown in Figure \ref{fig:critical_Bm_vs_n}, allowing to determine the critical field of the oblate--prolate droplet transition as evidenced by $\beta_2$ becoming positive. 
The critical field determined this way coincides nicely with the field predicted by numerically minimizing the energy of an presumably ellipsoidal droplet \cite{morozov_bifurcations_2000, erdmanis_magnetic_2017} as shown in Fig. \ref{fig:ba_vs_Bm}.
This energy minimization procedure, as well as experimental observations \cite{bacri_behavior_1994} also anticipate the droplet becoming oblate again at large enough magnetic fields $Bm\gtrsim 70$.
This reentrant transition is qualitatively observed in the short perturbed simulations with $\beta_2$ becoming negative at larger field strengths $Bm \gtrsim 50$ as shown in Figure \ref{fig:beta2_vs_Bm}. 
The discrepancy of the predicted field of the prolate--oblate transition may be explained by the fact that we started from an oblate shape and observed how the perturbations would grow or decay.
However to accurately capture the prolate--oblate transition, we would have to start from the prolate shape and observe its change.

\section{Conclusions and discussion} 
The elaboration of the numerical algorithm for 3D dynamics of magnetic droplet shapes allows to validate different relations describing their behavior and may be used for comparison with experimental data, thus providing information about the physical properties of the concentrated phase of magnetic colloids.
These colloids may possess very interesting properties due to their highly magnetic nature -- such as field dependent surface tension or dependence of rheological properties on the magnetic field.

The developed algorithm has been compared with some theoretical solutions where available -- equilibrium curves for droplet deformation in constant magnetic fields for various relative magnetic permeability $\mu$ values, the exponential decay of small elongations under surface tension, the dynamical behaviour around hysteresis ``bottleneck'' instability regions, as well as droplet dynamics in various rotating field configurations and the critical fields for oblate--prolate transition.
These comparisons allow to test the limits of various theoretical approximations widely used in description of magnetic fluid droplets, most notably the assumption of ellipsoidal shape.

The agreement of numerical results and theoretical consideratons also extends the limits of applicability of the simple magnetic fluid model description of these droplets which admittedly are a new kind of soft magnetic matter with a priori nontrivial mathematical representation.

It has also been shown to be applicable in prediction of full 3D droplet dynamics with arbitrary droplet--fluid viscosity ratios in uniform fields, both static and rotating up to moderately large droplet deformations. 
This is an important achievement, as experimentally magnetic fluid droplets are often not axisymmetric and can reach viscosity ratios of $\lambda\approx100$.
Our algorithm also allows for probing droplet dynamics at moderately fast rotating magnetic fields, where the fast--rotating field averaging approximation does not hold, which has not been previously possible.

The algorithm can capture characteristic behaviour of droplets in rotating magnetic fields, namely, following the external field at low frequencies and exhibiting a back-and-forth motion at moderately fast fields, similarly to a magnetic solid rods, as well as the droplet undergoing the oblate--prolate--oblate transition at rotating fields of increasing strength.

Important contribution of the present work is the numerical confirmation of the main aspects of the magnetic droplets behavior in static and rotating fields that opens the possibility of study of droplet dynamics in highly non--equilibrium situations not accessible at present by theoretical description.
Among the problems to be studied by the developed tools we should mention droplet dynamics in the intermediate range of frequencies, droplet breakup and ensembles of interacting magnetic droplets \cite{stikuts_spontaneous_2020}.

\section*{Acknowledgements}
A.L. acknowledges the financial support of the European Union’s Horizon 2020 research and innovation programme under grant agreement MAMI No. 766007. 

A.P.S. is thankful to SIA ``Mikrotīkls" and the Embassy of France in Latvia for supporting cotutelle studies.

A.P.S. acknowledges the financial support of ``Strengthening of the capacity of doctoral studies at the University of Latvia within the framework of the new doctoral model", identification No. 8.2.2.0/20/I/006

A.C. and A.P.S. acknowledge the financial support of grant of Scientific Council of Latvia lzp-2020/1-0149.

\section*{Declaration of Interests} The authors report no conflict of interest.

\appendix 

\section{Wielandt's deflation}
\label{app:Wielandt}
After straightforward algebraic operations with the regularized integrals in section \ref{sseq:regularization}, we get that the equation for velocity reads

\begin{equation}
\begin{split}
	v_k(\vect{y}) =& \frac{1}{8\pi}\int_S \left(    r_i n_i(\vect{x})n_k(\vect{y})  +  r_i n_i(\vect{y})n_k(\vect{x})  +  (1 - n_i(\vect{x}) n_i(\vect{y}) )r_k \vphantom{\frac{(num)}{(\vect{d}en)}} \right. \\
 &\qquad\qquad - \left. \frac{3 r_k  (  n_i(\vect{x}) + n_i(\vect{y}) ) r_i r_j n_j(\vect{x}) }{|\vect{r}|^2}    \right)\frac{dS_x}{|\vect{r}|^3} \\ 
	& +\frac{1}{8\pi}\int_S f_M(\vect{x}) n_i(\vect{x})G_{ik}(\vect{x},\vect{y})dS_x\\
	&+ \frac{1-\lambda}{8\pi}\int_S \left[v_i(\vect{x})-v_i(\vect{y})\right]T_{ijk}(\vect{x},\vect{y})n_j(\vect{x})dS_x \\
	& +v_k^\infty(\vect{y}) .
\end{split}
\end{equation}

To perform Wielandt's deflation, instead of solving for $\vect{v}$, we introduce an auxiliary field $\vect{w}$ similarly as \citep{pozrikidis_boundary_1992}, which is obtained from the integral equation that does not have the unwanted eigensolutions.

\begin{equation}
\begin{split}
	w_k(\vect{y}) =& \frac{1}{8\pi}\int_S \left(    r_i n_i(\vect{x})n_k(\vect{y})  +  r_i n_i(\vect{y})n_k(\vect{x})  +  (1 - n_i(\vect{x}) n_i(\vect{y}) )r_k \vphantom{\frac{(num)}{(\vect{d}en)}} \right. \\
 &\qquad\qquad - \left. \frac{3 r_k  (  n_i(\vect{x}) + n_i(\vect{y}) ) r_i r_j n_j(\vect{x}) }{|\vect{r}|^2}    \right)\frac{dS_x}{|\vect{r}|^3} \\ 
	& +\frac{1}{8\pi}\int_S f_M(\vect{x}) n_i(\vect{x})G_{ik}(\vect{x},\vect{y})dS_x\\
	&+ \frac{1-\lambda}{2} \left\{\frac{1}{4\pi}\int_S \left[w_i(\vect{x})-w_i(\vect{y})\right]T_{ijk}(\vect{x},\vect{y})n_j(\vect{x})dS_x \right. \\
	&\qquad\qquad\left. +w'_k(\vect{y}) - \frac{n_k(\vect{y})}{S} \int_S w_i(\vect{x}) n_i(\vect{x}) dS_x \vphantom{\frac{(num)}{(\vect{d}en)}} \right\}\\
	& +v_k^\infty(\vect{y}) ,
\end{split}
\end{equation}
where $S$ is the surface area of the droplet and $\vect{w}'$ is the projection of $\vect{w}$ on the motion of a rigid body obtained as \cite{zinchenko_novel_1997}
\begin{equation}
    \vect{w}'(\vect{y}) = \vect{V} + \vect{\Omega}\times(\vect{y}-\vect{y_c}),
\end{equation}
where $\vect{y_c}$ is the center of mass of the droplet, and $\vect{V}$ and $\vect{\Omega}$ are its translational and rotational velocities.
\begin{equation}
    \vect{y_c} = \frac{1}{S} \int_S \vect{x} dS_x
\end{equation}

\begin{equation}
    \vect{V} = \frac{1}{S} \int_S \vect{w}(\vect{x}) dS_x
\end{equation}

\begin{equation}
    \vect{\Omega} = \matr{M}^{-1} \cdot \int_S \left(\vect{\tilde{x}}\times \vect{w}(\vect{x}) \right)dS_x,
\end{equation}

\begin{equation}
    M_{ij} = \int_S \left(\tilde{x}_k\tilde{x}_k \delta_{ij} - \tilde{x}_i\tilde{x}_j  \right)dS_x,
\end{equation}

where $\vect{\tilde{x}} = \vect{x}-\vect{y_c}$. 

Finally the velocity is recovered by
\begin{equation}
    \vect{v}(\vect{y}) = \vect{w}(\vect{y}) - \frac{\kappa}{1+\kappa} \vect{w}'(\vect{y}) = \vect{w}(\vect{y}) - \frac{1-\lambda}{2} \vect{w}'(\vect{y}).
\end{equation}

\section{Normal field calculation without the tangential components} 

\label{app:normal_field_geometric}
Adjusting the approach seen in \cite{das_electrohydrodynamics_2017} we find an integral equation for the normal differences on the droplet surface
\begin{equation}
    \llbracket H_n (\vect{y}) \rrbracket \left( \frac{\mu}{\mu-1}-L(\vect{y}) \right) = H_{0n}(\vect{y}) - \int_S \left( \llbracket H_n(\vect{x}) \rrbracket - \llbracket H_n (\vect{y}) \rrbracket \right) \vect{n}(\vect{y}) \cdot \vect{\nabla}_y \frac{1}{4\pi r} dS_x,
\end{equation}
where we remember that $\vect{r} = \vect{y} - \vect{x}$ and where $\llbracket H_n (\vect{y}) \rrbracket =  H_n^{(e)} (\vect{y}) -  H_n^{(i)} (\vect{y})$, and we have
\begin{align}
    H^{(e)}_n = \frac{\llbracket H_n \rrbracket}{\mu-1} \mu \qquad H^{(i)}_n = \frac{\llbracket H_n \rrbracket}{\mu-1}.
\end{align}
 The function $ L(\vect{y})$ is determined uniquely by the droplet shape
\begin{equation}
    L(\vect{y}) = \frac{\vecf{n}{y}}{4\pi} \cdot \int_S \Big \{  [\vecf{n}{x}\cdot \vect\nabla \left(\frac{1}{r}\right)]\ [\vecf{n}{y}-\vecf{n}{x}] + \frac{\vect\nabla \cdot \vect{n} (\vect{x})}{r}\ \vecf{n}{x}  \Big \}\ dS_x.
\end{equation}

\section{Magnetic field tangential component}
\label{app:tangential_field}

We start with an expression from \cite{das_electrohydrodynamics_2017} where we replace the electric field $\vect{E}$ with its magnetic counterpart $\vect{H}$
\begin{equation}
    \vecf{H}{y}=\vecf{H_0}{y} - \int_S \llbracket H_n(\vect{x}) \rrbracket \nabl_y \frac{1}{4\pi r} \dd S_x - \frac{1}{2} \llbracket H_n(\vect{x}) \rrbracket \vecf{n}{y}.
\end{equation}

To get the tangential component, we take a cross product with $\vecf{n}{y}$:
\begin{equation}
    \vecf{n}{y}\times \vecf{H_t}{y}=\vecf{n}{y} \times \vecf{H_0}{y} + \frac{1}{4 \pi}\int_S  \llbracket H_n(\vect{x}) \rrbracket (\vecf{n}{y} \times \vect{r})   \frac{\dd S_x }{r^3}
    \label{eq:Ht_raw}
\end{equation}
where we remember that $\vect{r} = \vect{y} - \vect{x}$, also we note that only the tangential component of the field contributes to the cross product.

The integral, however, is strongly singular: as $\vect{x}\rightarrow \vect{y}$, the integrand scales as $O(1/r^2)$

Regularizing the equation by multiplying $\llbracket H_n(\vect{y}) \rrbracket$ with the identity
\begin{equation}
    \int_S \frac{\vecf{n}{x} \times \vect{r}}{4\pi r^3} \dd S_x = \int_V \nabl \times \left( \frac{\vec{r}}{4\pi r^3} \right) \dd V = \int_V \nabl \times \left( \nabl \frac{1}{4\pi r} \right) \dd V = 0
\end{equation}
and subtracting it from \eqref{eq:Ht_raw} we obtain
\begin{equation}
    \vecf{n}{y}\times \vecf{H_t}{y}=\vecf{n}{y} \times \vecf{H_0}{y} + \frac{1}{4 \pi}\int_S \frac{\dd S_x }{r^3} \left(\bigg[ \llbracket H_n(\vect{x}) \rrbracket \vecf{n}{y} - \llbracket H_n(\vect{y}) \rrbracket \vecf{n}{x} \bigg] \times \vect{r} \right)
    \label{eq:Ht_regularized}
\end{equation}
This integrand scales as $O(1/r)$ as $\vect{x}\rightarrow \vect{y}$ and can now be calculated using, for example, local polar coordinates centered at $\vect{y}$ for the singular elements.

The norm of the left hand side of \eqref{eq:Ht_regularized} simplifies to $|\vecf{n}{y}| |\vecf{H_t}{y}| \sin{(\pi/2)} = H_t(\vect{y})$, as the normal is unit length and it makes a right angle with the tangential component of the magnetic field. Therefore, we can calculate the magnitude of tangential magnetic field. This approach avoids the more numerically unstable method of numerical differentiation of the magnetic potential $\psi$ on the droplet surface.

\vspace{5mm}

\section{The virial theorem approach} 
\label{app:virial}

The droplet dynamics around this ``hysteresis jump'' region can be considered on the basis of the Rayleigh dissipation function $R$ when the energy of the droplet reads $dE/dt = -2R$ and $R$ is expressed as a quadratic function of the generalized velocity of the system -- in this case $de/dt$ or $d(a/b)/dt$:
\begin{equation}
    R = \frac{D\dot{e}^2}{2}.
\end{equation}
Combining the above relations in an Euler--Lagrange equation gives
\begin{equation}
\frac{\partial E}{\partial e} = -D\dot{e}.
\end{equation}

Near the threshold of the instability $\partial_e E = \partial^2_{ee} E = 0$ we have
\begin{equation}
    \frac{\partial E}{\partial e} \simeq \frac{1}{2}\frac{\partial^3 E}{\partial e^3}(e-e_c)^2 + H_c\ \frac{\partial^2 E}{\partial e \partial H}\  \frac{H-H_c}{H_c},
\end{equation}
or in terms of the constants $A,B,D$

\begin{equation}
    D\dot{e} = Ah + B(e-e_c)^2.
\label{eq:virial_approx}
\end{equation}

Since the viscosity of the magnetic droplet is much larger than the viscosity of surrounding liquid, it can be neglected. Then the condition of the force balance on the surface of the droplet reads
\begin{equation}
    -p + \sigma_{nn}^v = - \gamma \left( \frac{1}{R_1}+\frac{1}{R_2}\right) + \frac{\mu_0}{2} M_n^2,
\label{eq:virial_bc}
\end{equation}
with $\sigma^v$ being the viscous stress tensor and $M$ being the magnetization, and the equation of motion of the magnetic fluid assuming its ellipsoidal shape reads
\begin{equation}
    -\partial_i p + \partial_m \sigma_{im}^v = 0.
\label{eq:virial_eom}
\end{equation}

After multiplying \eqref{eq:virial_eom} by $x_k$ and integrating over the volume of the droplet and using the boundary condition \eqref{eq:virial_bc} gives the virial coefficients
\begin{equation}
    V_{ik} = \delta_{ik} \int p dV - \int \gamma x_k n_i\ \nabla \cdot  \vec{n}\ dS + \int x_k n_k \frac{\mu_0}{2} M_n^2 dS - \int \sigma_{ik}^v dV = 0
\end{equation}

Using $\int x_k n_i\ \nabla \cdot \vec{n}\ dS = - \int (\delta_{ik}-n_i n_k)\ dS$ among other relations \cite{cebers_virial_1985} it can shown that $V_{33}-\frac{1}{2}(V_{11} + V_{22})=0$ may be expressed as
\begin{equation}
\begin{gathered}
-\int \left[ \sigma_{33}^v - \frac{1}{2}(\sigma^v_{11} + \sigma^v_{22})\right] dV + \\
+ 2\pi\gamma R_0^2 \Bigg\{ \frac{\mu_0}{2}Bm \left[ \frac{(1-e^2)}{2}\left( \frac{(3-e^2)}{e^5} \log{\left( \frac{1+e}{1-e}\right)} - \frac{6}{e^4}\right) \right ] + \\ 
    + \frac{ \frac{(3-4e^2)\arcsin{e}}{e^3} -   \frac{(3-2e^2)(1-e^2)^{1/2}}{e^2}}{2(1-e^2)^{1/6}} \Bigg\} = 0,
\end{gathered}
\label{eq:virialdyn}
\end{equation}
with $Bm$ here being $\frac{M^2 R_0}{\gamma}$ rather than the usual $\frac{H^2 R}{\gamma}$. 

Remembering that droplet energy is expressed as

\begin{equation}
    E = -\frac{1}{2}\ \frac{\chi H^2}{1+\chi N / 4\pi}\  \frac{R_0^3}{3} + \frac{2\pi\gamma R_0^2}{(1-e^2)^{1/6}} \left[ \frac{\arcsin{e}}{e} + (1-e^2)^{1/2} \right],
\end{equation}
with $\chi$ being the magnetic susceptibility and $N$ the demagnetization coefficient, and a little algebra, we can 
identify the second term in \eqref{eq:virialdyn} to be equal to
\begin{equation}
    - \frac{\partial E}{\partial e}\ \frac{3(1-e^2)}{2e}.
\end{equation}

Moreover, in terms of Lagrangian displacement $\xi_3 = L_{33}x_3$ the first term in \eqref{eq:virialdyn} can be expressed as \begin{equation}
    -3\eta\dot{L}_{33} \frac{4\pi}{3}R_0^3.
\end{equation}

Finally, taking into account the relation \cite{blums_magnetic_1997}
\begin{equation}
    \dot{L}_{33} = \frac{2e\dot{e}}{3(1-e^2)},
\end{equation}
we obtain
\begin{equation}
     4\pi\eta R_0^3 \dot{e} + \frac{\partial E}{\partial e}\ \left( \frac{3(1-e^2)}{2e} \right)^2 = 0.
\end{equation}

Expressing this equation and the droplet energy in terms of an experimentally more convenient parameter -- the axis ratio $a/b = 1/\sqrt{1-e^2}$ (droplet energy further denoted by $\tilde{E}$ and axis ratio by $e$), allows one to identify the capillary relaxation time $\tau_c = \eta R_0 / \gamma$ in the dynamical equation
\begin{equation}
    \frac{d}{dt}\left( \frac{a}{b} \right) =  -\frac{1}{\tau_c}\ \frac{9 (1-e^2)^{1/2}}{8e}\  \frac{\partial \tilde{E}}{\partial e}.
\label{eq:virial_ab_dynamics}
\end{equation}

Evaluation of the expansion of $\frac{\partial \tilde{E}}{\partial e}$ at either $e_c$ or $H_c$ allows to obtain expressions for the constants $A$ or $B$ accordingly. The constant $D$ can be easily identified after integration of \eqref{eq:virial_ab_dynamics} and reproducing \eqref{eq:tange} for small $t$:
\begin{equation}
    \left(\frac{a}{b}\right) - \left(\frac{a}{b}\right)_c = - \frac{A}{2D}\ \frac{t}{\tau_c} \left( \frac{H^2}{H_c^2} - 1 \right).
\end{equation}

The constants obtained in this manner yield \eqref{eq:ABD}:
\begin{equation*}
    A =  \frac{8 \pi}{3}\ Bm_c\ \frac{6e_c + (e_c^2-3) \log \left[ \frac{1+e_c}{1-e_c}\right]}{2e_c^4},\ \ \ \
    B =  -\frac{1}{2} \frac{\partial^3 \tilde{E}}{\partial e^3} \frac{(1-e_c^2)^3}{e_c^2}, \ \ \ 
    D = \frac{8 e_c}{9 (1-e_c^2)^{1/2}}.
\end{equation*}

\bibliographystyle{elsarticle-num}
\bibliography{biblio}

\end{document}